\documentclass[10pt,aps,prb,twocolumn,noshowpacs]{revtex4-1}
\usepackage{amsmath,graphicx,amsbsy,amssymb,amsmath,epsfig,latexsym,wasysym,pifont,mathrsfs,natbib}
\usepackage{float} \newfloat{widefig}{thp}{lop} \usepackage{comment}
\usepackage{array, makecell} \usepackage{verbatim} \usepackage{datetime}
\usepackage{multirow,array} \usepackage{hyperref} \usepackage{color} \usepackage{setspace}
\usepackage{subcaption}
\usepackage{graphicx}
\usepackage{dirtytalk}
\usepackage[british]{babel}
\usepackage[utf8,latin1]{inputenc}
\UseRawInputEncoding
\usepackage{csquotes}
\usepackage{comment}
\DeclareMathOperator*{\dprime}{\prime \prime}
\usepackage[export]{adjustbox}

\hypersetup{colorlinks,
  linkcolor=blue,%
  citecolor=blue,%
  urlcolor=blue}

\begin{document}

\title{\textbf{Symmetry Lowering Through Surface Engineering and Improved Thermoelectric Properties in MXenes}}

\author{Himanshu Murari}
\email[]{hmurari@iitg.ac.in}
\affiliation{Department of Physics, Indian Institute of Technology
  Guwahati, Guwahati-781039, Assam, India.}  
\author{Subhradip Ghosh}
\email{subhra@iitg.ac.in} \affiliation{Department of Physics,
  Indian Institute of Technology Guwahati, Guwahati-781039, Assam,
    India.} 
\begin{abstract}  
 Despite ample evidence of their influences on the transport properties of two-dimensional solids, the interrelations of reduced symmetry, electronic and thermal transport, have rarely being discussed in the context of thermoelectric materials. With the motivation to design new thermoelectric materials with improved properties, we have addressed these by performing first-principles Density Functional Theory based calculations in conjunction with semi-classical Boltzmann transport theory on a number of compounds in the MXene family. The symmetry lowering in parent M$_{2}$CO$_{2}$ MXenes are done by replacing transition metal $M$ on one surface, resulting in Janus compounds MM$^{\prime}$CO$_{2}$. Our calculations show that the thermoelectric figure-of-merit can be improved significantly by such surface engineering. We discuss in detail, both qualitatively and quantitatively, the origin behind high thermoelectric parameters for these compounds. Our in-depth analysis shows that the modifications in the electronic band structures and degree of anharmonicity driven by the dispersions in the bond strengths due to lowering of symmetry, an artefact of surface engineering, are the factors behind the trends in the thermoelectric parameters of the MXenes considered. The results also substantiate that the compositional flexibility offered by the MXene family of compounds can generate complex interplay of symmetry, electronic structure, bond strengths and anharmonicity which can be exploited to engineer thermoelectric materials with improved properties.     
\end{abstract}


\maketitle

\section{Introduction\label{intro}}
Thermoelectric materials can produce electricity directly from waste heat without the use of any mechanical effort. These materials are viable sources for renewable and green energy \cite{sootsman2009new}. However, the real challenge lies in improvement of their conversion efficiencies.  A dimensionless quantity called the figure of merit (ZT) is used to assess the efficiency of thermoelectric (TE) materials. The figure of merit is given as $ZT = \frac{S^{2}\sigma}{(\kappa_{e}+\kappa_{l})}$, where $S$ is the Seebeck coefficient, $\sigma$, $\kappa_{e}$, and $\kappa_{l}$ are the electrical conductivity, electronic and lattice thermal conductivity \cite{disalvo1999thermoelectric}, respectively. In oder to achieve a  conversion efficiency of $\sim$ 30\%, equal to that of a Carnot refrigerator, $ZT$ should be $\sim$ 3. Till date, the conversion efficiency of the most advanced, commercially available TE material  is only $\sim 10-12$\%. This is due to the fact that the transport coefficients $S,\sigma$ and $\kappa_{e}$ are interrelated ($S \propto \sigma^{-1}, \sigma \propto \kappa_{e}$) and hence they cannot be tuned independently. $\kappa_{l}$ is the sole independent parameter. Consequently, much effort has been devoted to find suitable strategies to reduce $\kappa_{l}$ and thus improve $ZT$\cite{yu2019ultralow, biswas2012high}. Reducing the dimension of the system is one of the contemporary methods in this direction \cite{hicks1996experimental, hicks1993effect}. The resulting quantum confinement effect  plays a significant role in tweaking the densities of states (DOS), giving rise to Van-Hove singularity, that enhances the thermopower $S$ \cite{dresselhaus2007new}. On top of that,  phonons are constrained in low-dimension materials leading to to enhanced phonon-phonon scattering and reduced $\kappa_{l}$\cite{alam2013review}. Among low-dimensional materials, two-dimensional (2D) ones have found to have excellent TE efficiency.

Since the discovery of graphene in 2004\cite{novoselov2004electric}a number of novel 2D materials\cite{li2020recent} have been identified and thoroughly investigated physically and conceptually for different applications. MXenes, a unique class of 2D transition metal carbides and nitrides was discovered in 2011\cite{naguib2011two}. MXenes are the 2D derivatives of ternary transition metal carbides/nitrides, the MAX compounds, by selective etching of the A layer \cite{hu2015mo}. MXenes have the chemical formula of M$_{n+1}$X$_{n}$T$_{x}$ (n=1-3), where M is transition metal, X is carbon/nitrogen, and T is the functional group passivating the surfaces. The compositional flexibility of MXenes has made them useful for multitude of applications\cite{lukatskaya2013cation, xie2013extraordinarily,shahzad2016electromagnetic}. Subsequently, their thermoelectric properties were investigated. Thermoelectric properties of Ti$_{3}$C$_{2}$, the MXene that was  discovered first has been most extensively investigated due to its multiple functional properties, exhibited a high $\sigma$ in film form \cite{filmti3c2}. However, first-principles simulation predicted maximum $ZT$ for Ti$_{3}$C$_{2}$ flakes to be only $0.11$ \cite{flaketi3c2}. Few experimental investigations on Mo$_{2}$CT$_{x}$ \cite{kim2017thermoelectric} and Nb$_{2}$CT$_{x}$ \cite{nanogen} reported reasonable values of thermoelectric power factor ($S^{2}\sigma$). Various first-principles simulations on a number of different MXenes investigated different aspects of structure-property relationships and provided useful information about their thermoelectric properties. Large $S$ and $\kappa_{l}$ smaller than the reference 2D materials like Graphene and MoS$_{2}$ were obtained from computations on Sc$_{2}$CT$_{x}$ MXenes \cite{sc2c}. Calculations on Ti$_{2}$CT$_{x}$ MXene showed that inspite of larger specific heat and group velocity, the phonon relaxation time lowered the $\kappa_{l}$ significantly in Ti$_{2}$CO$_{2}$ \cite{pccp2018}. Simulations with Y$_{2}$CH$_{2}$ MXenes without inclusion of $\kappa_{l}$ showed a maximum $ZT$ close to 1 \cite{y2ct2}. A comparative study of Ti$_{2}$CO$_{2}$, Zr$_{2}$CO$_{2}$ and Hf$_{2}$CO$_{2}$ by first-principles simulations \cite{gandi2016thermoelectric} including the lattice effects attributed the comparatively lower $\kappa_{l}$ (and higher $ZT$)in Ti$_{2}$CO$_{2}$ to the increase in the Umklapp scattering involving acoustic and optical phonon branches. The maximum $ZT$, however, was only 0.45. Investigations into the effects of structure on the thermoelectric properties for these three systems was carried out by first-principles simulations with the same level of approximation \cite{sarikurt2018influence}. It was inferred that symmetry lowering can improve $ZT$ drastically with major influence coming from lowering of $\kappa_{l}$. 

Lowering of symmetry in MXenes can be easily accomplished by changing the chemical composition of the transition metal surfaces. To this end, thermoelectric measurements and simulations were performed on a few MXenes. A large Power Factor in Mo$_{2}$TiC$_{2}$T$_{x}$ MXene was experimentally obtained at room temperature \cite{nanogen} enabling it to be used as a thermoelectric nano-generator. A Power Factor, higher by two order of magnitudes than Ti$_{3}$C$_{2}$T$_{x}$ films, was obtained for Mo$_{2}$Ti$_{2}$C$_{3}$T$_{x}$ at 800 K \cite{kim2017thermoelectric} raising the prospect of obtaining a high $ZT$. First-principles based simulations predicted high $ZT$ values $\sim$ 3 for p-type Ti$_{2}$MoCF$_{2}$ \cite{tanusri} and Cr$_{2}$TiC$_{2}$(OH)$_{2}$ MXenes \cite{jing2019superior}. In both cases the major factor was very low value of $\kappa_{l}$; the reason attributed to increased phonon-phonon anharmonic scattering. In case of M$_{2}$X MXenes, manipulation of transition metal surfaces through such substitution, would lead to Janus MXenes with chemical formula MM$^{\prime}$X. Janus 2D compounds are obtained by breaking the inversion symmetry upon replacing the constituents of one of the two chemically identical surfaces. The Janus counterparts of 2D dichalcogenides have turned out to exhibit superior functional properties \cite{lu2017janus, deng2019enhanced} as seen from various experiments. With regard to TE properties, Janus compounds  WSX(X=Te,Se) were found to possess much better properties than dichalcogenide WS$_{2}$ \cite{patel2020high}. However, till date there is only one first-principles based simulation results available for Janus MXenes. A large Power factor in comparison to MoS$_{2}$, Mo$_{2}$TiC$_{2}$ and Mo$_{2}$Ti$_{2}$C$_{3}$ were obtained for n-TiMoCO$_{2}$ Janus compound \cite{wong2020high}. Based on the trends in the electrical and electronic TE parameters of this compound, it was predicted that lack of inversion symmetry may lower $\kappa_{l}$ and elevate $ZT$. 

Although the results obtained for various MXenes suggest that lowering the symmetry would lead to an improved $ZT$ by substantial reduction of $\kappa_{l}$ due to increased anharmonicity, driven by the breaking of the inversion symmetry, the results of Reference \cite{jing2019superior} imply that huge reduction in $\kappa_{l}$ for -OH functionalised Cr$_{2}$TiC$_{2}$, in comparison to -O and -F functionalised ones, was obtained even though there was no change in symmetry. Moreover, a recent study on Ta$_{2}$CS$_{2}$ MXene \cite{ta2cs2} showed that the breaking of inversion symmetry promotes $\kappa_{l}$. This was attributed to the weakening of the anharmonic scattering whose origin was in the re-distribution of bond strengths due to lowering of symmetry. These indicate a connection between symmetry, strengths of chemical bonds and anharmonic scattering, an aspect that has not been addressed in depth. 

In this paper, we perform first-principles based investigations to address this connection by computing the TE parameters of Ti$_{2}$CO$_{2}$, Mo$_{2}$CO$_{2}$, Zr$_{2}$CO$_{2}$ and Hf$_{2}$CO$_{2}$ MXenes and five Janus compounds TiMoCO$_{2}$, ZrMoCO$_{2}$, HfMoCO$_{2}$, TiZrCO$_{2}$  and TiHfCO$_{2}$. Our motivation is two-fold: (1) to explore the possibility of obtaining significantly high $ZT$ upon breaking inversion symmetry through surface engineering in MXenes and (2) to provide deeper insights into the inter-relations between symmetry lowering, electronic structure, lattice dynamics and various transport parameters with special emphasis on harmonic and anharmonic components in phonon contributions. By computing every conceivable quantities to understand the trends in the TE parameters, we demonstrate that surface manipulation of M$_{2}$CO$_{2}$ MXenes to form Janus compounds enhances anharmonic scattering significantly, the root of which can be traced back to dispersions in the intra-surface bond strengths between a pair of constituents, resulting in significant lowering of $\kappa_{l}$ and elevation in $ZT$ to the desired magnitude.  

\section{Computational detail}
The  first-principles calculations are performed by Density Functional Theory (DFT)\cite{hohenberg1964inhomogeneous,kohn1965self} method as implemented in the Vienna ab initio simulation package (VASP)\cite{kresse1996efficient}. The Projected Augmented Wave (PAW) \cite{blochl1994projector} pseudopotentials are used for the calculations. The exchange-correlation part of the Hamiltonian is approximated by  the Perdew-Burke-Ernzerhof (PBE) functional \cite{perdew1996generalized} under generalized gradient approximation (GGA). The structural optimisations are carried out with a kinetic energy cutoff of 550 eV for the plane-wave basis set. The Brillouin zone (BZ) integration is performed using the Monkhorst-Pack scheme\cite{monkhorst1976special} with a grid size of 16$\times$16$\times$1. The convergence criteria for total energy and Hellmann-Feynman forces are kept at  $10^{-7}$eV and  $10^{-3}$eV $\AA^{-1}$, respectively. A denser $k$-mesh of 36$\times$36$\times$1 grid is chosen for calculation of the densities of states. To avoid the interaction among periodic images along the $c$ direction, a vacuum of $20 \AA$ is set up. Since the Janus compounds considered here are not yet synthesised experimentally, Ab initio molecular dynamics (AIMD) simulations are performed to ascertain their thermodynamic stabilities. The AIMD simulations are performed using a canonical ensemble (NVT) in the Nos\'e-Hoover heat bath\cite{nose1984unified}.  A 3$\times$3$\times$1 supercell with 4$\times$4$\times$1 $k$-point grid are used for the calculations. The simulations are run for 20 ps. 

To obtain the electronic transport properties (Seebeck coefficient ($S$), electrical conductivity ($\sigma$/$\tau$), and electronic thermal conductivity ($\kappa_{e}$/$\tau$)) we have solved the  Boltzmann transport equations  under constant relaxation time approximation (CRTA) and rigid band approximation (RBA), as implemented in the BoltzTrap2 package\cite{madsen2018boltztrap2}. In the CRTA approach, electronic relaxation time ($\tau$) is held constant while evaluating the transport coefficients. In the RBA, the electronic structure of the material is invariant with the change of temperature and carrier density; only the Fermi energy shifts. Since the transport coefficients obtained this way are $\tau$ dependent, we have used the deformation potential (DP) theory\cite{bardeen1950deformation} to evaluate $\tau$. The expressions for electrical conductivity tensor($\sigma_{\alpha\beta}(T,\mu)$),electronic thermal conductivity tensor($\kappa^{o}_{\alpha\beta}(T,\mu)$), and Seebeck coefficient tensor($S_{\alpha\beta}(T,\mu)$) are:
\begin{equation}\label{sig}
	\sigma_{\alpha\beta}(T,\mu)= \frac{1}{\Omega}\int {\bar{\sigma}_{\alpha\beta}(\epsilon)}\Big{[}-{\frac{\partial f(T,\epsilon,\mu)}{\partial\epsilon}}\Big{]}d\epsilon
\end{equation},
\begin{equation}\label{kappae}
	\kappa^{o}_{\alpha\beta}(T,\mu)=\frac{1}{e^{2}T\Omega}\int{\bar{\sigma}_{\alpha\beta}(\epsilon)}(\epsilon - \mu)^{2}\Big{[}-{\frac{\partial f(T,\epsilon,\mu)}{\partial\epsilon}}\Big{]}d\epsilon
\end{equation}

\begin{equation}\label{seebeck}
	S_{\alpha\beta}(T,\mu)=\frac{1}{eT\Omega\sigma_{\alpha\beta}(T,\mu)}\int{\bar{\sigma}_{\alpha\beta}(\epsilon)}(\epsilon-\mu)\Big{[}-{\frac{\partial f(T,\epsilon,\mu)}{\partial\epsilon}}\Big{]}d\epsilon
\end{equation}
where
\begin{equation}\label{bar_sigma}
	{\bar{\sigma}}_{\alpha\beta}(\epsilon) = \frac{e^{2}}{N}\sum_{i,k}{\tau v_{\alpha}(i,k) v_{\beta}(i,k)\delta(\epsilon-\epsilon_{i,k})}
\end{equation}
and
\begin{equation}\label{v}
	v_{\alpha}(i,k)=\frac{1}{\hbar}\frac{\partial\epsilon_{i,k}}{\partial k_{\alpha}}
\end{equation}

$\alpha$ and $\beta$ are the tensor indices, $\Omega$ is the volume of the unit cell, $\mu$ is the chemical potential, and $f$ is the Fermi-Dirac distribution function. $N$ denotes the number of $k$ points sampled, and $e$  the electron charge. $v_{\alpha}(i,k)$($\alpha=x,y,z$) indicates the $\alpha$-th component of the group velocity of the carriers, corresponding to the $i$-th energy band. Equations (\ref{sig})-(\ref{v}) suggest that the transport coefficients can be evaluated from the group velocity $v_{\alpha}(i,k)$ obtained from the band structure computed with DFT.

The vibrational properties are obtained by finite displacement method as implemented in the Phonopy package\cite{togo2015first}. 4$\times$4$\times$1 supercell with 6$\times$6$\times$1 $k$-point grid have been used to obtain the harmonic force constants and phonon dispersions.The lattice thermal conductivities have been calculated by solving the phonon Boltzmann transport equation using an iterative method implemented in ShengBTE code\cite{li2014shengbte}. The lattice thermal conductivity is given by,
\begin{equation}
    \kappa^{\alpha\beta}_{l} = \frac{1}{k_{B}T^{2}VN}\sum_{\lambda}f_{0}(f_{0}+1)(\hbar\omega_{\lambda})^{2}v_{\lambda}^{\alpha}F_{\lambda}^{\beta}
    \label{kl_eq}
\end{equation}
where $V$ is the volume of the unit cell, and $N$ the number of $q$ points sampled in the Brillouin zone.  $\omega_{\lambda}$ and $v_{\lambda}$ are the angular frequency and group velocity of the phonon branch $\lambda$, respectively. $f_{0}$ is the Bose distribution function and $F_{\lambda}^{\beta}=\tau_{\lambda}^{0}(v_{\lambda}+\Delta_{\lambda})$ is the form of linearised BTE when only two and three phonon scatterings are considered. $\tau_{\lambda}^{0}$ is the relaxation time of phonon mode $\lambda$, the inverse of which is equal to the sum of all possible transition probabilities between phonon modes $\lambda$, $\lambda^{\prime}$ and $\lambda^{\dprime}$,
\begin{equation}
    \frac{1}{\tau_{\lambda}^{0}} = \frac{1}{N}\Big{(}\sum_{\lambda^{\prime}\lambda^{\dprime}}^{+}\Gamma_{\lambda\lambda^{\prime}\lambda^{\dprime}}^{+} + \sum_{\lambda^{\prime}\lambda^{\dprime}}^{-}\frac{1}{2}\Gamma_{\lambda\lambda^{\prime}\lambda^{\dprime}}^{-} + \sum_{\lambda^{\prime}}\Gamma_{\lambda\lambda^{\prime}}\Big{)}
\end{equation}
The first two terms in the above equation correspond to the three-phonon scattering rates. The first one describes absorption and the second term  the emission processes. The third term corresponds to isotopic scattering. The expressions for absorption (+) and emission (-) processes are given as
\begin{equation}
    \Gamma_{\lambda\lambda^{\prime}\lambda^{\dprime}}^{+} = \frac{\hbar\pi}{4}\frac{f_{0}^{\prime}-f_{0}^{\dprime}}{\omega_{\lambda}\omega_{\lambda^{\prime}}\omega_{\lambda^{\dprime}}}|\phi_{\lambda\lambda^{\prime}\lambda^{\dprime}}^{+}|^{2}\Delta(\omega_{\lambda}+\omega_{\lambda^{\prime}}-\omega_{\lambda^{\dprime}})
\end{equation}
\begin{equation}
    \Gamma_{\lambda\lambda^{\prime}\lambda^{\dprime}}^{-} = \frac{\hbar\pi}{4}\frac{f_{0}^{\prime}+f_{0}^{\dprime}+1}{\omega_{\lambda}\omega_{\lambda^{\prime}}\omega_{\lambda^{\dprime}}}|\phi_{\lambda\lambda^{\prime}\lambda^{\dprime}}^{-}|^{2}\Delta(\omega_{\lambda}-\omega_{\lambda^{\prime}}-\omega_{\lambda^{\dprime}})
\end{equation}
where $|\phi_{\lambda\lambda^{\prime}\lambda^{\dprime}}^{\pm}|$ is the scattering matrix elements and which depends on the anharmonic IFCs ($\Phi_{ijk}^{\alpha\beta\gamma}$) as,
\begin{equation}
    |\phi_{\lambda\lambda^{\prime}\lambda^{\dprime}}^{\pm}| = \sum_{i\epsilon u.c.} \sum_{j,k} \sum_{\alpha\beta\gamma} \Phi_{ijk}^{\alpha\beta\gamma}\frac{e_{\lambda}^{\alpha}(i) e_{\lambda^{\prime}}^{\beta}(j) e_{\lambda^{\dprime}}^{\gamma}(k)}{\sqrt{M_{i}M_{j}M_{k}}}
    \label{eq5}
\end{equation}
 $M_{i}$ is the mass of $i^{th}$ atom,  $e_{\lambda}^{\alpha}(i)$ denotes the $\alpha$-th component of an eigenvector of mode $\lambda$ of $i$-th atom. In this case, summation over $i$ denotes the space within the unit cell, whereas $j$ and $k$ span the entire system.

For evaluating the third-order IFCs, a supercell of 3$\times$3$\times$1 with eight nearest neighbours is considered. The  harmonic and anharmonic IFCs are used to compute the lattice thermal conductivity $\kappa_{l}$ and solve BTE for phonons. For calculating $\kappa_{l}$, a $q$-grid of 24$\times$24$\times$1 is used. The $\kappa_{l}$ thus obtained is scaled by a factor $c/z$ , where $c$ is vacuum height, and $z$ the thickness of the material considered. For this work, thickness is taken to be the distance between the top and bottom $O$ planes. The convergence of $\kappa_{l}$ with respect to sizes of the supercell and the $q$-grid  is checked by separate calculations using supercells of size 3$\times$3$\times$1 and 4$\times$4$\times$1 and $q$-grids of size 24$\times$24$\times$1 and 26$\times$26$\times$1. The difference is found to be about 1\% only.

\section{Results and discussion}

\subsection{Structural Parameters, Bond Strengths and Band Structures}
The ground state structural parameters of M$_{2}$CO$_{2}$  and MM$^{\prime}$CO$_{2}$, (M, M$^{\prime}$)=Ti,Mo,Zr,Hf), are presented in Table \ref{tab1}. The ground state structures are obtained from DFT calculations by calculating total energies corresponding to structural models that differ from each other depending upon the sites of passivation by -O. The models along with their total energies are shown in Figures 1,2 and Table 1, supplementary material. We find that the sites of passivation on the M and M$^{\prime}$ surfaces in the Janus compounds remain unchanged from those in their M$_{2}$C counterparts. The calculated lattice constants are in excellent agreement with the available results. Expectedly, the lattice constants of Janus MM$^{\prime}$CO$_{2}$ MXenes are in between those of end point MXenes M$_{2}$CO$_{2}$ and M$^{\prime}_{2}$CO$_{2}$. Since the inversion symmetry of the M$_{2}$CO$_{2}$ MXenes is broken in Janus MXenes (the space group of M$_{2}$CO$_{2}$ compounds is  $P\bar{3}m1$ while that of Janus MM$^{\prime}$CO$_{2}$ is $P3m1$), there is significant fluctuations in the M/M$^{\prime}$-O/C  bond lengths, as is evident from Table \ref{tab1}. For Janus compounds with Mo as M$^{\prime}$ element, Mo-C(M-C) and Mo-O(M-C) bonds increase (decrease) slightly in comparison to those in parent Mo$_{2}$CO$_{2}$(M$_{2}$CO$_{2}$) compounds. However, the common feature across parent and Janus compounds is that the M/M$^{\prime}$ -O bonds are shorter than the M/M$^{\prime}$-C bonds. The fluctuations in the bond lengths lead to fluctuations in the bond strengths resulting in anharmonicity in the system having an effect on the lattice thermal conductivity $\kappa_{l}$.
      
\vspace{0.5mm}
 \begin{table}[H]
	\centering
	\resizebox{1.0\columnwidth}{!}{%
		\begin{tabular}{|c| c|c| c| c|c| c| c|}
			\hline
			MXenes & GS-Model & a & E$_{g}$  & d$_{M-C}$ & d$_{M-O}$ & d$_{M'-C}$ & d$_{M'-O}$  \\
                   &    & (\AA) & (eV) & (\AA) & (\AA) & (\AA) & (\AA) )\\ 
			\hline
			$Ti_{2}CO_{2}$& HH &3.03 (3.03)\cite{gandi2016thermoelectric}& 0.32 & 2.18  & 1.97   & - & -  \\
			$Mo_{2}CO_{2}$& CC &2.89(2.88)\cite{khazaei2014two} & -    & 2.15  & 2.06   & - & - \\
			$Zr_{2}CO_{2}$& HH &3.31(3.26)\cite{gandi2016thermoelectric} & 0.97 & 2.36  & 2.12   & - & - \\
			$Hf_{2}CO_{2}$& HH &3.26(3.26)\cite{gandi2016thermoelectric} & 1.03 & 2.33  & 2.10   & - & - \\
			\hline	
			$TiMoCO_{2}$& HC &2.95 & 0.16 & 2.15 & 1.95 & 2.13 & 2.08  \\
			$ZrMoCO_{2}$& HC &3.10 & 0.18 & 2.29 & 2.07 & 2.18 & 2.13 \\
			$HfMoCO_{2}$& HC &3.08 & 0.29 & 2.26 & 2.05 & 2.18 & 2.12 \\
			$TiZrCO_{2}$& HH &3.18 & 0.74 & 2.25 & 2.01 & 2.31 & 2.08 \\
			$TiHfCO_{2}$& HH &3.16 & 0.72 & 2.24 & 2.00 & 2.28 & 2.07 \\
			\hline
		\end{tabular}
  }
		\caption{Structural parameters and electronic band gaps  for the systems considered.}
		\label{tab1}
	\end{table}
 \begin{figure*}
    \centering
    \includegraphics[width=0.80\linewidth,center]{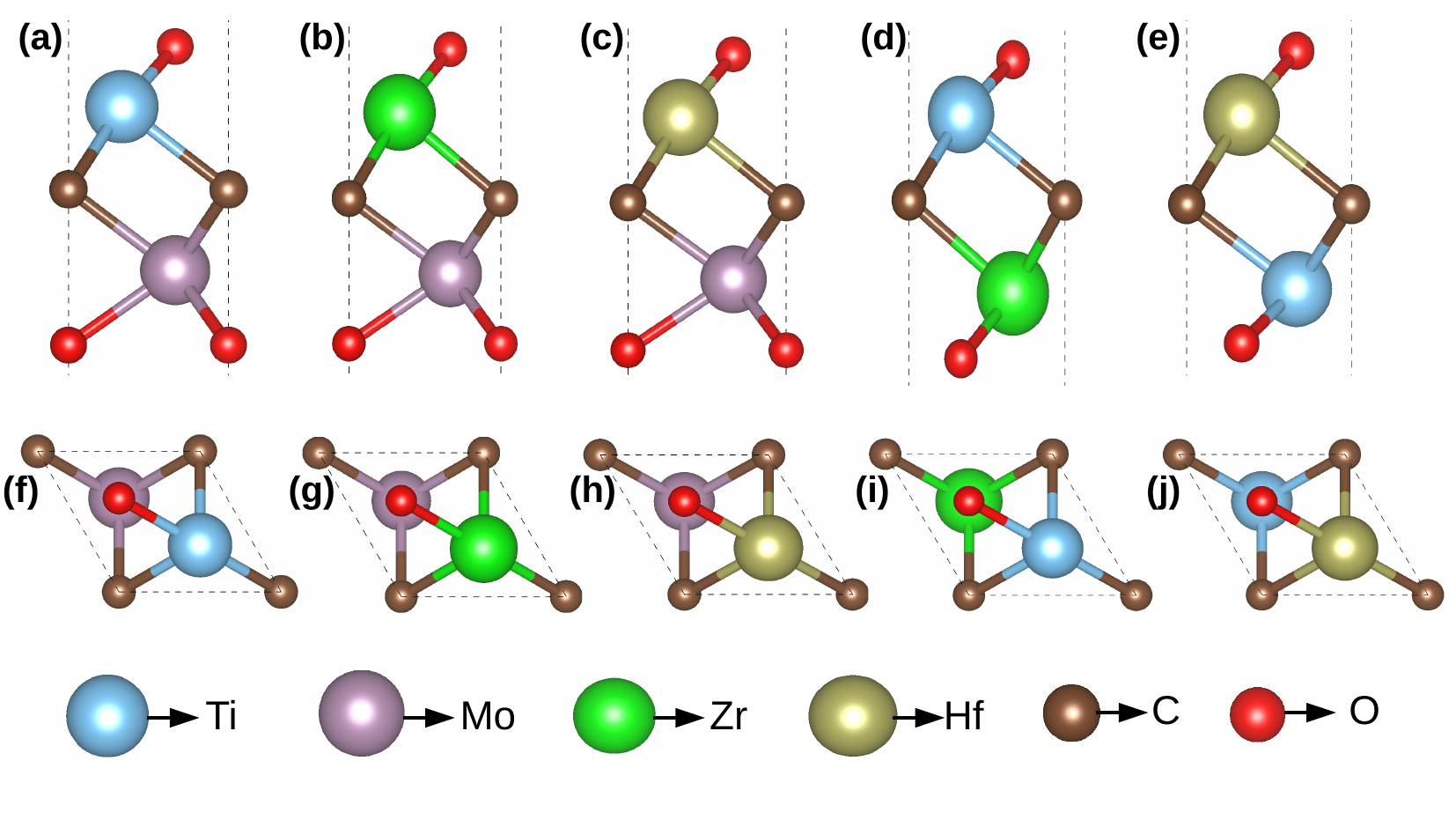}
    \caption{Ground state structure of Janus MXenes, (a)-(e) side view, (g)-(k) top view}
    \label{gs}
\end{figure*}
An assessment of the bond strengths and it's correlation with the dispersions in the M/M$^{\prime}$-C/O bond lengths is done by the Crystal Orbital Hamilton Population (COHP) method \cite{deringer2011crystal} as implemented in LOBSTER code\cite{maintz2016lobster}. In this method, a measure of the bond strengths are obtained from the energy integrated COHP, that is, from the energy averaged bond-weighted densities of states between a pair of atoms. In Table \ref{tab2}, the strengths between various bonds in parent and Janus compounds are shown. We find that the bond strengths are clearly correlated with the variations in the bond lengths. Among the parent M$_{2}$CO$_{2}$ MXenes, M-C and M-O bond strengths are comparable in case of Mo$_{2}$CO$_{2}$. For the other three compounds, M-O bonds are significantly stronger than the M-C bonds. For TiZrCO$_{2}$ and TiHfCO$_{2}$ Janus compounds, Ti-O and Ti-C bonds weaken considerably while M$^{\prime}$-O and M$^{\prime}$-C bonds strengthen, in comparison to those for their respective parent M$_{2}$CO$_{2}$ MXenes. For MMoCO$_{2}$ Janus, Mo-C and Mo-O bonds weaken only slightly while M-O and M-C bonds strengthen with respect to the respective bond strengths in the corresponding parent compounds.
\begin{table}[H]
    \centering
	\resizebox{0.75\columnwidth}{!}{%
    \begin{tabular}{|c |c |c |c |c |}
    \hline
     \multirow{2}{*}{MXenes}  & \multicolumn{4}{|c|}{Bond strength (eV)}  \\
     \cline{2-5}
    & M-C & M-O & M$^{'}$-C & M$^{'}$-O \\
    \hline
    Ti$_{2}$CO$_{2}$ & -2.85 & -3.88 & - & - \\
    \hline
    Mo$_{2}$CO$_{2}$ & -3.42 & -3.56 & - & - \\
    \hline
    Zr$_{2}$CO$_{2}$ & -3.15 & -4.14 & - & - \\
    \hline
    Hf$_{2}$CO$_{2}$ & -2.99 & -4.18 & - & - \\
    \hline
    TiMoCO$_{2}$ & -2.91 & -4.01 & -3.37 & -3.41 \\
    \hline
    ZrMoCO$_{2}$ & -3.37 & -4.53 & -3.36 & -3.10 \\
    \hline
    HfMoCO$_{2}$ & -3.37 & -4.51 & -3.31 & -3.16 \\
    \hline
    TiZrCO$_{2}$ & -2.46 & -3.50 & -3.34 & -4.39 \\
    \hline
    TiHfCO$_{2}$ & -2.48 & -3.56 & -3.31 & -4.37 \\
    \hline
    \end{tabular}
}
    \caption{Bond strengths for parent (M$_{2}$CO$_{2}$) and Janus (MM${'}$CO$_{2}$) MXenes obtained from COHP analysis.}
    \label{tab2}
\end{table}
\begin{figure}
    \centering
    \includegraphics[width=1.0\linewidth]{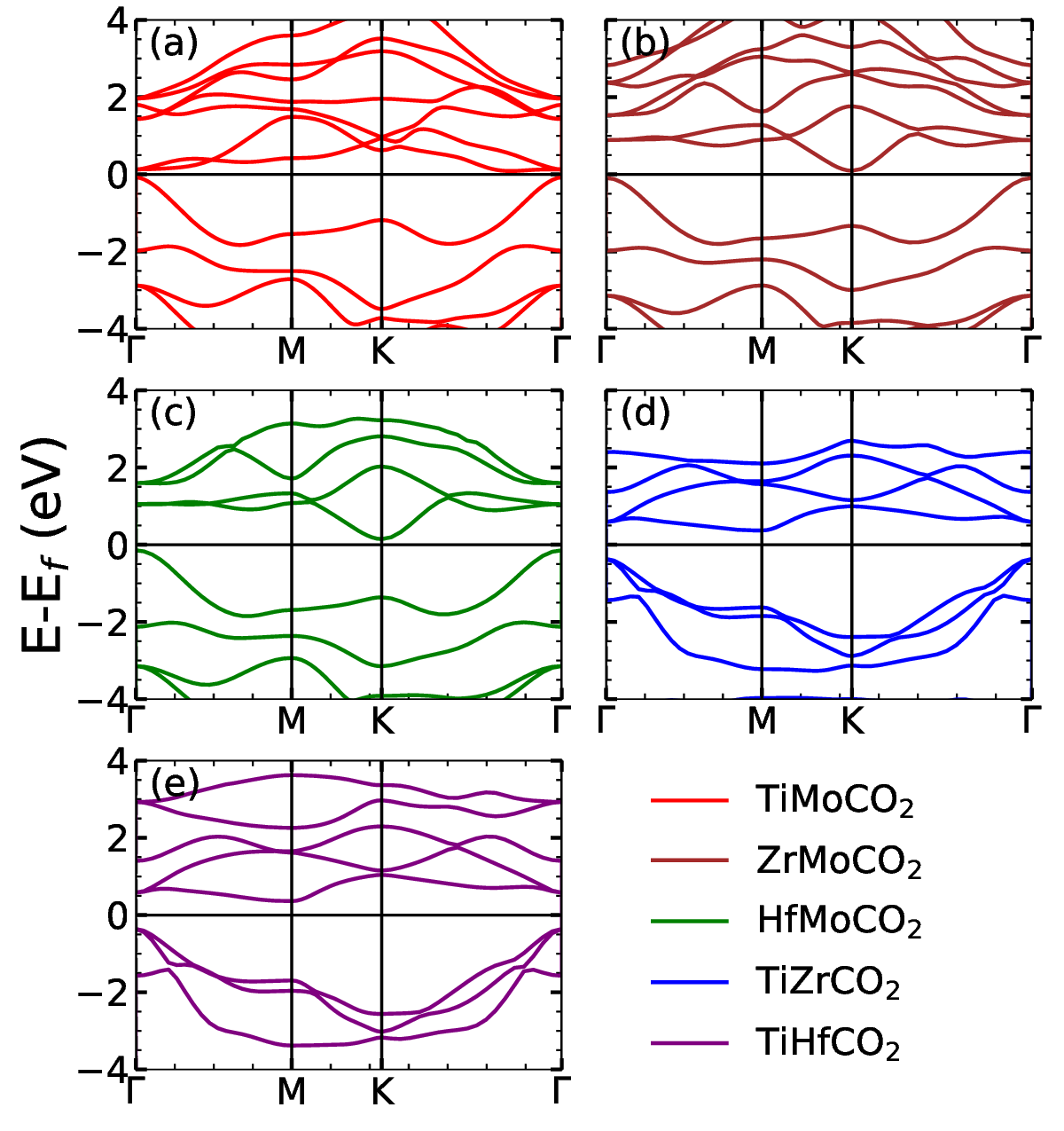}
    \caption{Electronic band structures of the  Janus MXenes considered.}
    \label{jbs}
\end{figure}
The band structures of Janus and the parent MXenes are shown in Figure \ref{jbs} and Figure 3, supplementary material, respectively. Densities of states of the Janus MXenes are shown in Figure 4, supplementary material. In agreement with existing results, we find that except Mo$_{2}$CO$_{2}$ which is a semi-metal, all M$_{2}$CO$_{2}$ MXenes considered are semiconductors. However all MM$^{\prime}$CO$_{2}$ Janus compounds, even with Mo as M$^{\prime}$ component, are semiconductors. The semiconducting nature of Mo-based Janus MXenes are due to the presence of the other transition metal constituent. The electronic structures in Figure 4, supplementary material, clearly show that in the anti-bonding part of the spectra, hybridisations between $d$ orbitals of Mo and the other transition metals that are located higher in energy, open the gaps in Mo-based Janus compounds. The magnitude of the gap depends on the position of the conduction band of M element in MM$^{\prime}$CO$_{2}$ MXene. The band gaps of TiZrCO$_{2}$ and TiHfCO$_{2}$ decrease considerably in comparison with Zr$_{2}$CO$_{2}$ and Hf$_{2}$CO$_{2}$, respectively (Table \ref{tab1}). Once again the reductions are due to the positions of Ti states that are closer to Fermi levels. Analysing the band structures we find that for ZrMoCO${_2}$ and HfMoCO${_2}$, the valence band maxima (VBM) and conduction band minima (CBM) are located at the $\Gamma$ and K points, respectively. In contrast, they are located at $\Gamma$ and M points, respectively, for TiZrCO$_{2}$ and TiHfCO$_{2}$. The electronic structure of TiMoCO$_{2}$ has very distinctive features.  The VBM and CBM are located at $\Gamma$ and along K-$\Gamma$ direction,respectively. The band structure shows a flat band close to Fermi level in the conduction region, the presence of which is reflected in the large Van Hove singularity in the densities of states.  Presence of such flat band near Fermi level would have profound impact on electronic transport properties as the effective mass of different carriers will be significantly different.
\subsection{Phonon spectra and dynamical stability}
\begin{figure}[H]
\captionsetup{font={small},skip=0.25\baselineskip}
\captionsetup[subfigure]{font={bf,small},skip=-1pt, singlelinecheck=false}
	\begin{subfigure}{0.49\linewidth}
		\centering
		\vspace{0.01mm}
		\includegraphics[width=1\linewidth]{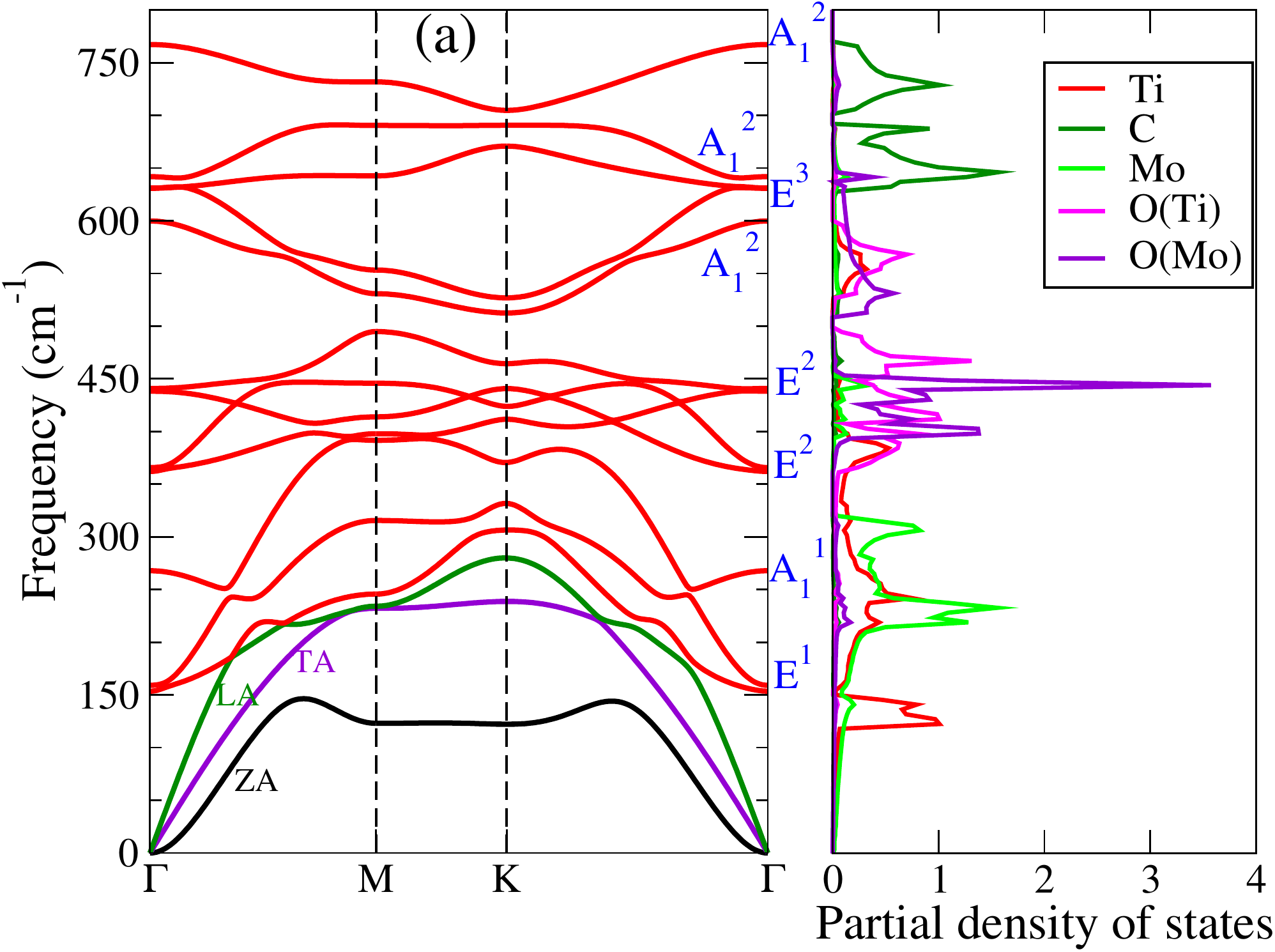}
		\label{ph_timo}
	\end{subfigure}
	\hfill
	\begin{subfigure}{0.49\linewidth}
		\centering
		\includegraphics[width=1\linewidth]{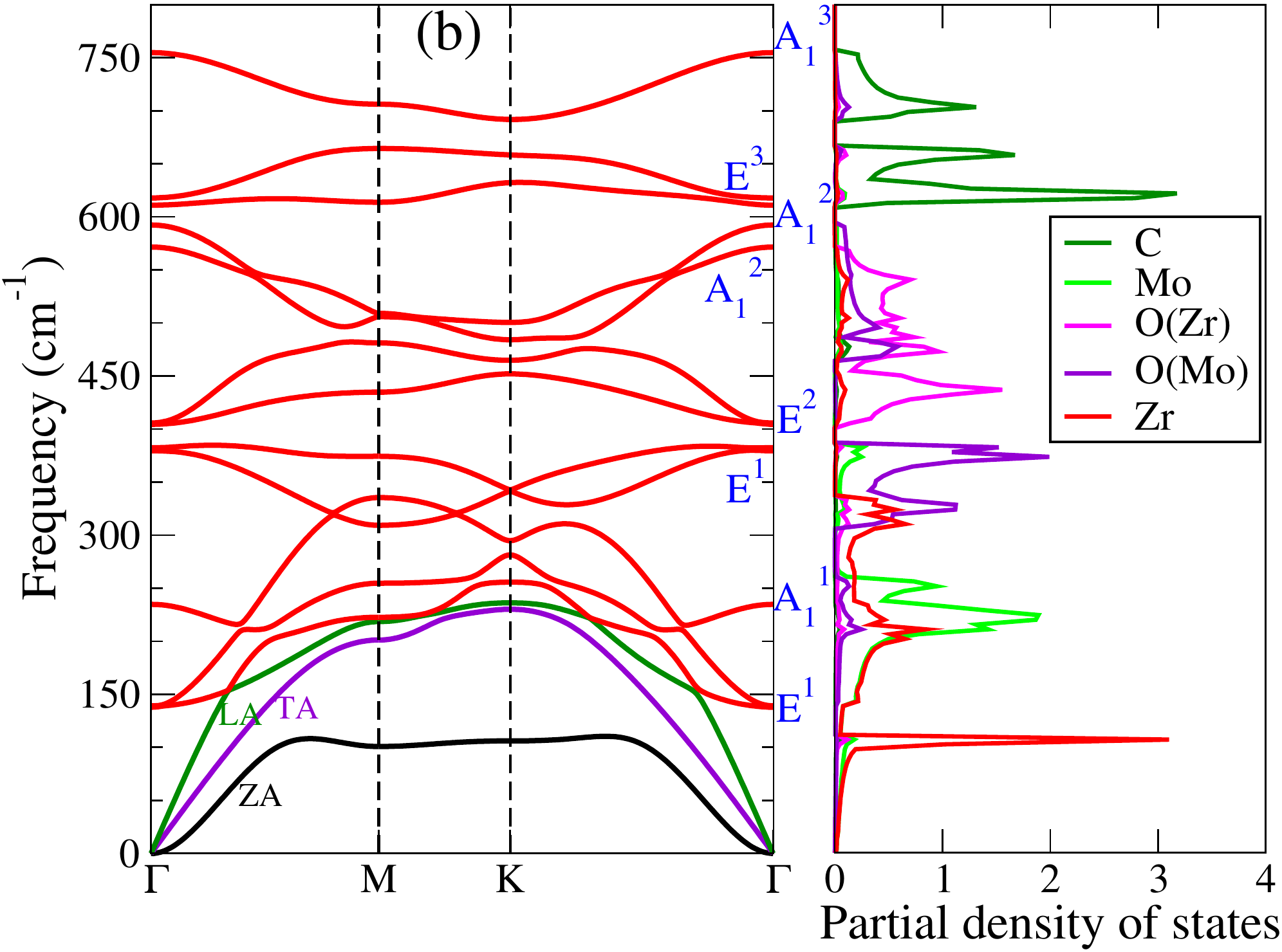}
		\label{ph_zrmo}
	\end{subfigure}
	\hfill
	\begin{subfigure}{0.49\linewidth}
		\centering
		\includegraphics[width=1\linewidth]{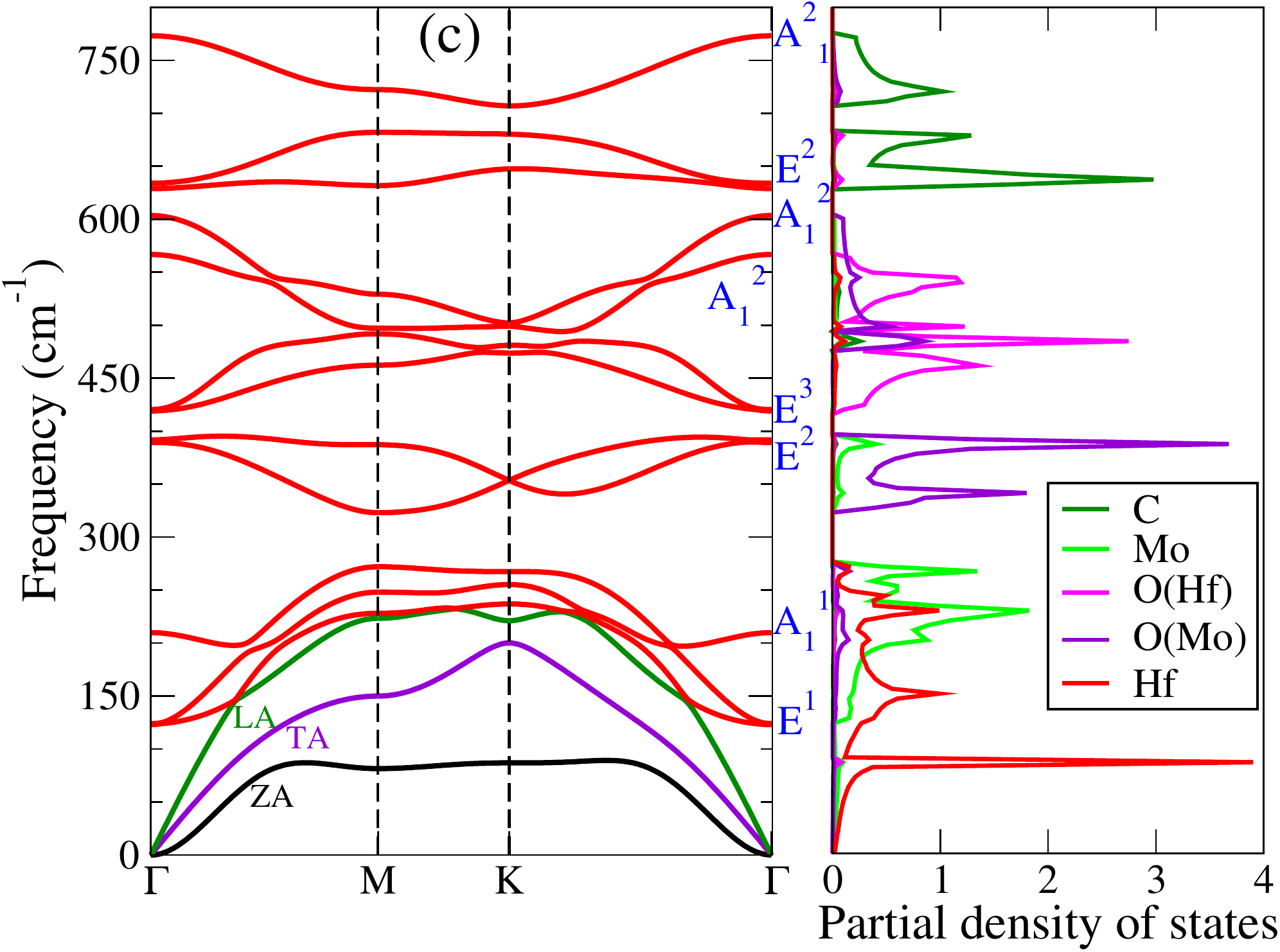}
		\label{ph_hfmo}
	\end{subfigure}
	\hfill
	\begin{subfigure}{0.49\linewidth}
		\centering
		\includegraphics[width=1\linewidth]{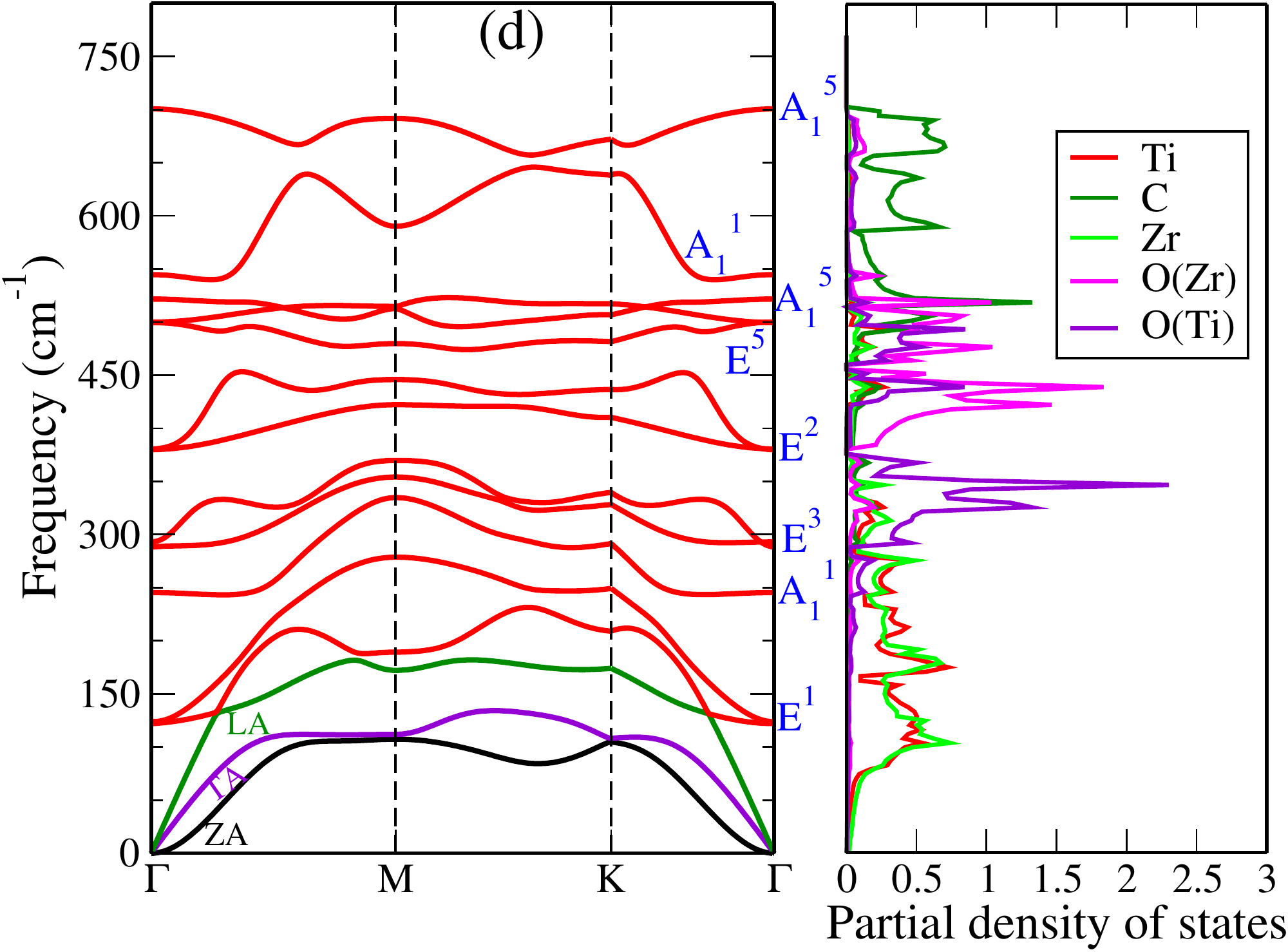}
		\label{ph_tizr}
	\end{subfigure}
 	\begin{subfigure}{0.49\linewidth}
		\centering
		\includegraphics[width=1\linewidth]{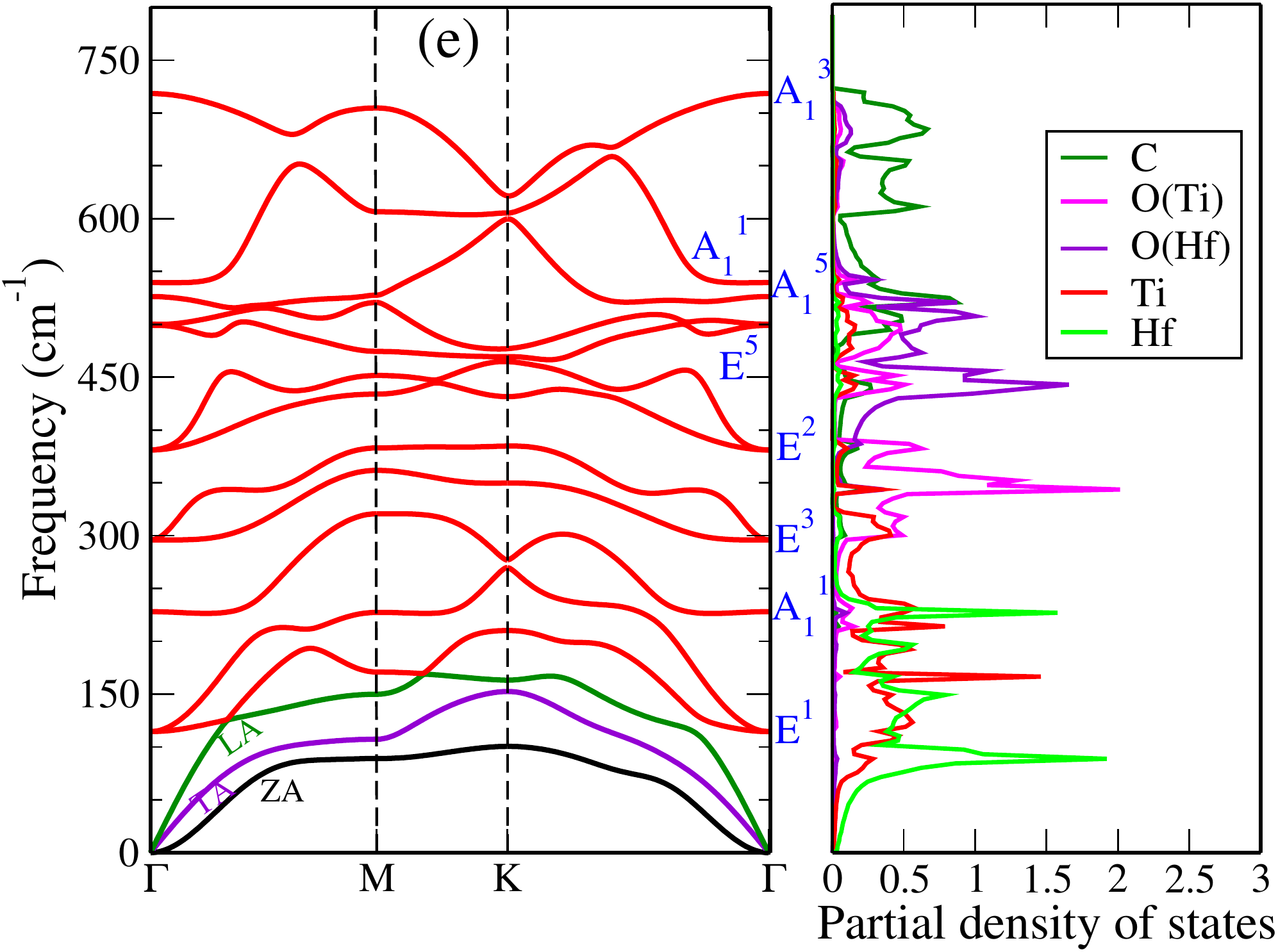}
		\label{ph_tihf}
	\end{subfigure}
        \hspace{5mm}
         \begin{subfigure}{0.38\linewidth}
		\includegraphics[width=1\linewidth]{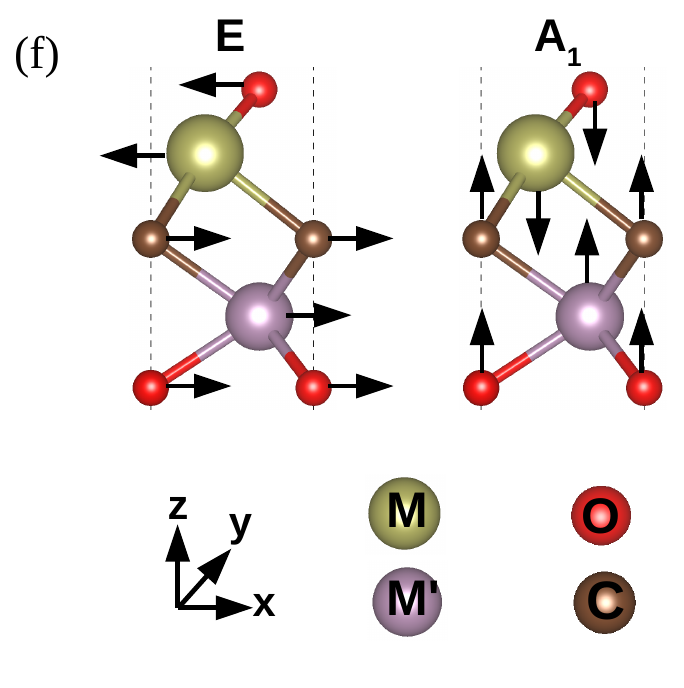}
		\label{ph_tihf}
	\end{subfigure}
	\caption{(a)-(e) Phonon dispersion and phonon density of states of the Janus MXenes considered. (f) The schematic diagram for vibration modes}
\label{phonon}
\end{figure}

Dynamical stabilities of the Janus compounds are assessed by computing their phonon spectra. The  phonon dispersion curves and the phonon partial density of states for MM$^\prime$CO$_{2}$ (M$_{2}$CO$_{2}$) Janus (parent) MXenes are shown in Figure \ref{phonon} (Figure 5, supplementary material). 

We find all systems to be dynamically stable in their respective ground states. For M$_{2}$CO$_{2}$ MXenes, the heavier elements (Ti, Mo, Zr, and Hf) dominate the low-frequency (acoustic and lower optical modes) phonon modes, while O and the lightest C atoms dominate the mid and high-frequency optical modes (Figure 5, supplementary materials). Similarly, in Janus MXenes, M and M$^{\prime}$ atoms majorly contribute in the lower frequency range (Figure \ref{phonon}). Mid-frequency optical modes show significant hybridization between two O atoms attached to M and M$^{\prime}$ surfaces. C atoms are responsible for the high frequency vibrations. We have performed the group theory analysis to identify the Raman and infrared (IR) active modes for Janus(for parent MXenes also; refer to Figure 6, supplementary material) MXenes. According to group theory, these vibrations can be expressed as irreducible representations ($\Gamma(vib)$) of C$_{3v}$ point group and can be represented as 
\begin{equation}
   \Gamma(vib) = 4A_{1}+4E
\end{equation}
where $A_{1}$ refers out-of-plane vibrations, and $E$ denotes the in-plane vibrations, that are doubly degenerate at the $\Gamma$ point. For this point group, all modes are Raman and IR active. Vibrational patterns for the  modes are schematically shown in Figure \ref{phonon}(f) (Vibration patterns corresponding to all $E$ and $A_{1}$ modes are shown in Figure 6, supplementary material). 
\section{Transport properties}
In this section we discuss the electronic and thermal transport properties of M$_{2}$CO$_{2}$ and MM$^{\prime}$CO$_{2}$ MXenes. By making a comparative assessment of Janus MXenes and the corresponding end point compounds, we analyse the effects of inversion symmetry breaking on the transport properties.
\subsection{Electronic transport parameters}
\begin{figure}[H]
	\centering
	\includegraphics[width=1.0\linewidth]{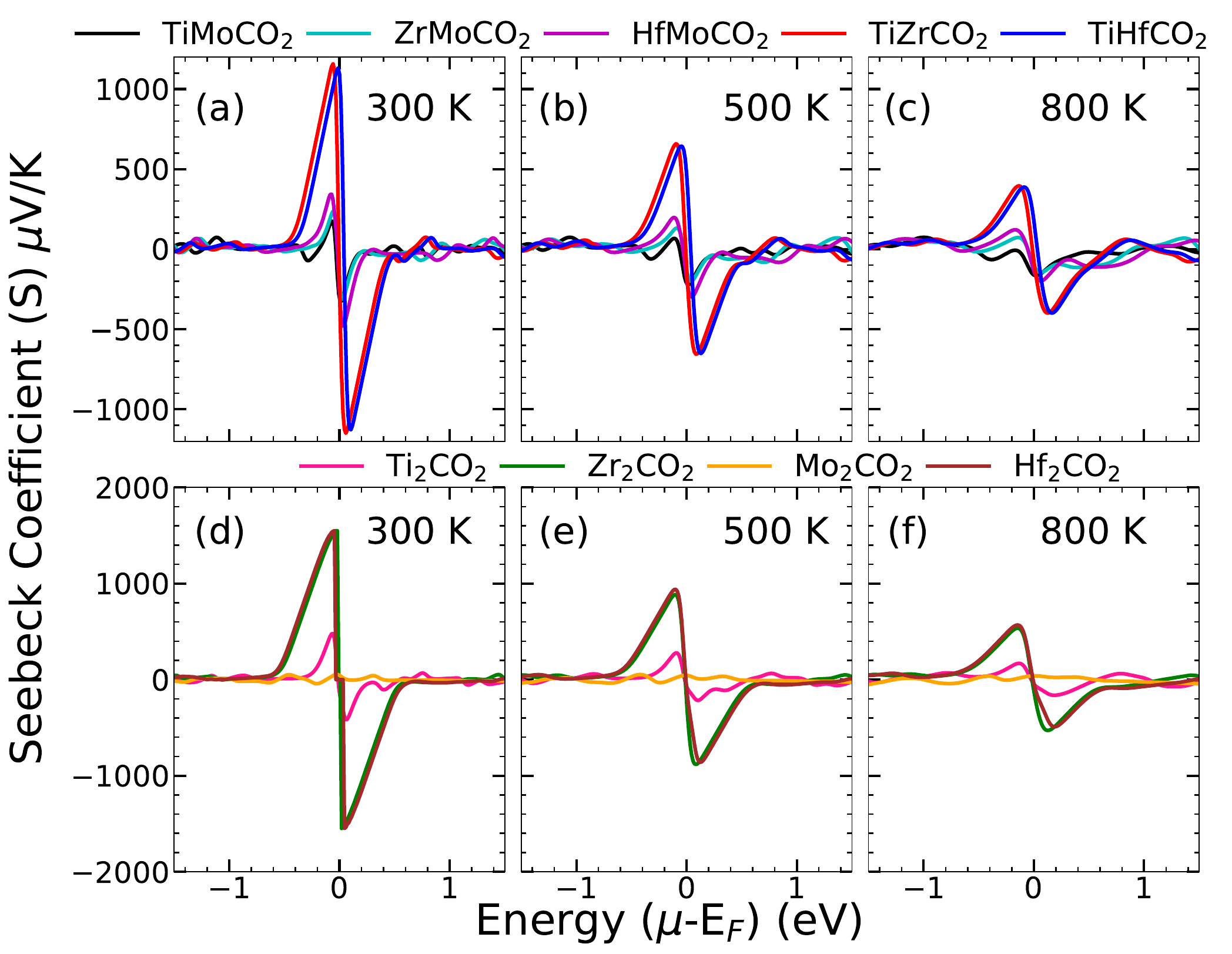}
	\caption{Seebeck coefficient ($S (\mu V/K$)) as a function of energy at different temperatures. The top(bottom) panel shows results for Janus(Parent) MXenes.}
	\label{sb}
\end{figure}

\begin{figure}[H]
	\centering
	\includegraphics[width=1.0\linewidth]{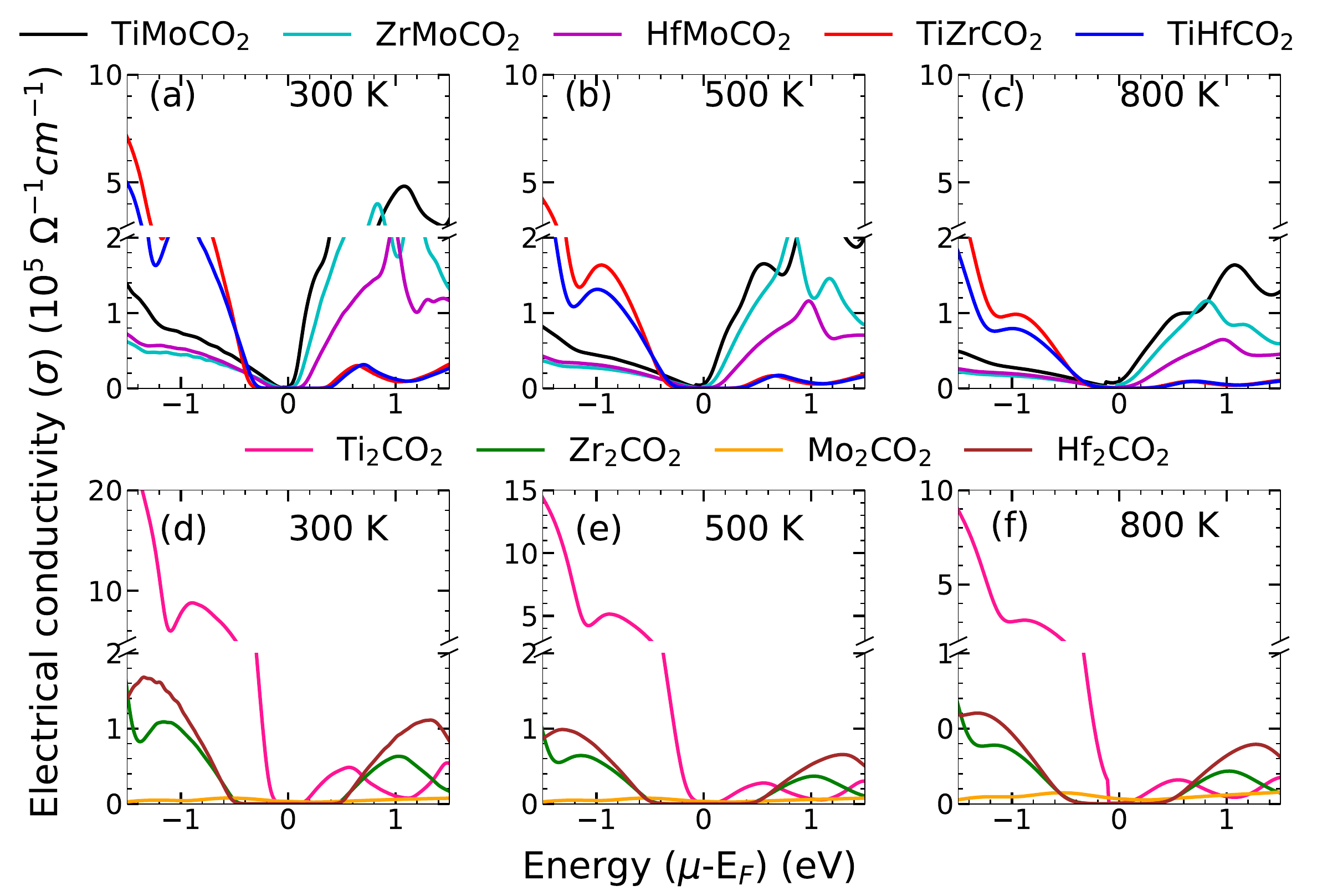}
	\caption{Electrical conductivity ($\sigma$ ($\Omega^{-1}cm^{-1}$)) as a function of energy at different temperatures. The top(bottom) panel shows results for Janus(Parent) MXenes.}
	\label{sigma}
\end{figure}

In Figures \ref{sb}, \ref{sigma} and  \ref{kappae}, we show results for variations of Seebeck coefficient ($S$), electrical conductivity ($\sigma$) and electronic part of thermal conductivity ($\kappa_{e}$) with energy, at three different temperatures, respectively. While Equations (\ref{sig})-(\ref{v}) solved under CRTA and RBA provide $S, \sigma/\tau$ and $\kappa_{e}/\tau$, $\tau$ is calculated from DP theory. 
The results suggest that TiZrCO$_{2}$ and TiHfCO$_{2}$ Janus have significantly higher values of S($\sim$1000$\mu$V/K) in comparison to the ones containing Mo($\sim$250$\mu$V/K). Among the parent MXenes, Hf$_{2}$CO$_{2}$ and Zr$_{2}$CO$_{2}$ have large values of S($\sim$1600$\mu$V/K) in comparison with the other two. Our results for parent MXenes agree well with existing results \cite{gandi2016thermoelectric,sarikurt2018influence,wong2020high,khazaei2014two}. Moreover, the calculated values of $S$ for the Janus MXenes are either higher or are comparable to those for the established thermoelectric materials \cite{shu2023high,yu2019ultralow}. 

The comparative behaviour of the electrical conductivity ($\sigma$), on the other hand, is different for different charge carriers. For p-type carrier, Ti$_{2}$CO$_{2}$ has significantly large $\sigma$ in comparison to Zr$_{2}$CO$_{2}$ and Hf$_{2}$CO$_{2}$, which have comparable values.  For n-type carrier, $\sigma$ is highest for Hf$_{2}$CO$_{2}$, followed by Zr$_{2}$CO$_{2}$ and Ti$_{2}$CO$_{2}$. However, all three have comparable values of $\sigma$.  Mo$_{2}$CO$_{2}$ has almost negligible electrical conductivity in comparison to the rest of the parent MXenes, irrespective of the type of charge carrier. In case of Janus compounds too, different trends are observed for different charge carriers. For p-type carrier, Mo based Janus compounds have substantially lower values of $\sigma$ in comparison to TiZrCO$_{2}$ and TiHfCO$_{2}$ which have almost identical magnitudes. For n-type carrier, the trend is exactly opposite. Like $S$, maximum  $\sigma$ for Janus compounds are noticeably smaller in comparison with that of the parent MXenes. The behaviour of electronic thermal conductivity $\kappa_{e}$ follows the trends of $\sigma$. This is consistent with Widemann-Franz relationship $\kappa_{e} \propto \sigma$.

The trends in the electronic transport parameters $S$ and $\sigma$ can be qualitatively understood from the trends of the effective mass of the carriers, $m^{*}_{h}$ (holes) and $m^{*}_{e}$ (electrons), the band gaps $E_{g}$, the mobilities $\mu_{h}$ and $\mu_{e}$ of electrons and holes, respectively. The effective masses and the mobilities of the carriers for parent and Janus MXenes considered in this work, are computed using DP theory and are presented in Table 2, supplementary material. For parabolic bands and energy independent scattering $S \propto \frac{m^{*}}{n^{2/3}}$ where $m^{*}$ is the effective mass of the carrier and $n$ the carrier density. Considering the parent compounds first, we find that $m^{*}_{h, Ti_{2}CO_{2}}(m^{*}_{e, Ti_{2}CO_{2}}) < (>) m^{*}_{h, Zr_{2}CO_{2}} (m^{*}_{e, Zr_{2}CO_{2}}) \approx (>) m^{*}_{h, Hf_{2}CO_{2}} (m^{*}_{e, Hf_{2}CO_{2}})$. Since band gap of Ti$_{2}$CO$_{2}$ is substantially smaller than that of the other two, it has a larger electron density $n$. As a result $S$ for Zr$_{2}$CO$_{2}$ and Hf$_{2}$CO$_{2}$ are much larger than Ti$_{2}$CO$_{2}$ irrespective of the carrier type. The reason for quantitatively smaller (larger) $S$ for Janus TiZrCO$_{2}$ and TiHfCO$_{2}$ in comparison with Zr$_{2}$CO$_{2}$, Hf$_{2}$CO$_{2}$ (Ti$_{2}$CO$_{2}$) can also be understood in a similar way. The effective masses of Mo-based Janus compounds, irrespective of charge carrier type, are larger than those of the three parent MXenes considered. The band gaps follow exactly the opposite trend implying that the carrier densities of Janus are larger. However, $S$ of Janus ZrMoCO$_{2}$ and HfMoCO$_{2}$ are significantly smaller in comparison to $S$ of Zr$_{2}$CO$_{2}$ and Hf$_{2}$CO$_{2}$. Reduction of such magnitude is probably due to substantial reduction of band gaps in Janus compounds which has a larger effect on $n$, superseding the effect of m$^{*}$. A comparison between $S$ of TiMoCO$_{2}$ and Ti$_{2}$CO$_{2}$ shows that $S_{TiMoCO_{2}} \gtrsim S_{Ti_{2}CO_{2}}$. This is because of two reasons: (a) the reduction in the band gap in TiMoCO$_{2}$ in comparison with Ti$_{2}$CO$_{2}$ is not as large as is the case of the other two sets of Janus-parent MXenes and (b) the increase in the effective mass in TiMoCO$_{2}$ in comparison with Ti$_{2}$CO$_{2}$ is much larger with respect to the other two sets. Such increase in effective mass in TiMoCO$_{2}$ is due to flatter bands in comparison to the other systems.

The comparative magnitudes of $S$ for the five Janus MXenes can also be understood by comparing the effective masses of holes and electronic band gaps. From the results presented in Table \ref{tab1} and Table 2, supplementary material, we find that $m^{*}_{h, TiMoCO_{2}} < m^{*}_{h, ZrMoCO_{2}} < m^{*}_{h,HfMoCO_{2}} > m*_{h,TiZrCO_{2}}=m^{*}_{h,TiHfCO_{2}}$ and $E_{g,TiMoCO_{2}}< E_{g,ZrMoCO_{2}} < E_{g,HfMoCO_{2}} << E_{g,TiZrCO_{2}} \approx E_{g,TiHfCO_{2}}$. As a result $n_{TiMoCO_{2}}>n_{ZrMoCO_{2}} > n_{HfMoCO_{2}} >> n_{TiZrCO_{2}} \approx n_{TiHfCO_{2}}$. The large electron densities in the Mo-based Janus compounds as compared with TiZrCO$_{2}$ and TiHfCO$_{2}$ nullifies the effect of larger $m^{*}_{h}$ in Mo-based Janus MXenes. Consequently, the Mo-based Janus MXenes have much smaller Seebeck coefficient as compared to the rest.

The electrical conductivity $\sigma \propto \mu n$. For the parent and Janus MXenes considered in this work, we find that the trends in $\sigma$ are largely dictated by $\mu$ of the carriers. In case of parent Mxenes, the hole mobilities are related as $\mu^{h}_{Ti_{2}CO_{2}} >> \mu^{h}_{Zr_{2}CO_{2}} \approx \mu^{h}_{Hf_{2}CO_{2}}$ while the electron mobilities follow $\mu^{e}_{Hf_{2}CO_{2}} > \mu^{e}_{Zr_{2}CO_{2}} \approx \mu^{e}_{Ti_{2}CO_{2}}$. The corresponding $\sigma$ for these compounds exactly follow these trends. Mo$_{2}$CO$_{2}$ is not considered in this context as it has an extremely low $\sigma$ and it being a semi-metal, mobilities associated with different charge carriers is irrelevant. For Janus compounds, the hole mobilities follow the trend $\mu^{h}_{TiZrCO_{2}} > \mu^{h}_{TiHfCO_{2}} >> \mu^{h}_{TiMOCO_{2}} > \mu^{h}_{ZrMoCO_{2}} \approx \mu^{h}_{HfMoCO_{2}}$ while the electron mobilities are related as $\mu^{e}_{ZrMoCO_{2}} > \mu^{e}_{HfMoCO_{2}} >> \mu^{e}_{TiMoCO_{2}} > \mu^{e}_{TiZrCO_{2}} \approx \mu^{e}_{TiHfCO_{2}}$. For the hole carriers, trend of $\sigma$ exactly follows the trend in carrier mobility. For the electron carriers, $\sigma$ of Mo-based Janus compounds are significantly higher than the remaining two. This cannot be explained by the trend in $\mu^{e}$ alone. The Mo-based Janus have much lower band gaps compared to the other two leading to very high carrier density $n$ resulting in substantially high $\sigma$. Even among the three Mo-based Janus, the qualitative trend of $\sigma$ for electron carriers can be explained only if both $\mu^{e}$ and $n$ are considered. The mobility turns out to be the deciding factor too for explaining the comparative trends between parent and Janus MXenes. For hole carriers, $\mu^{h}_{Zr_{2}CO_{2}} << \mu^{h}_{TiZrCO_{2}} << \mu^{h}_{Ti_{2}CO_{2}}$ and $\sigma_{Zr_{2}CO_{2}} < \sigma_{TiZrCO_{2}} < \sigma_{Ti_{2}CO_{2}}$. Same behaviour is observed for TiHfCO$_{2}$. For Mo-based Janus MXenes, hole mobilities are much smaller than those of Ti$_{2}$CO$_{2}$, TiZrCO$_{2}$ and TiHfCO$_{2}$. This explains the reason behind significantly smaller $\sigma$ of Mo-based Janus compounds in comparison with the other MXenes considered. For electron carriers, electrical conductivity of TiZrCO$_{2}$ and TiHfCO$_{2}$ are not very different from that of the corresponding parent MXenes. This is consistent with close values of their electron mobilities. 
\begin{figure}[H]
\centering
\includegraphics[width=1.0\linewidth]{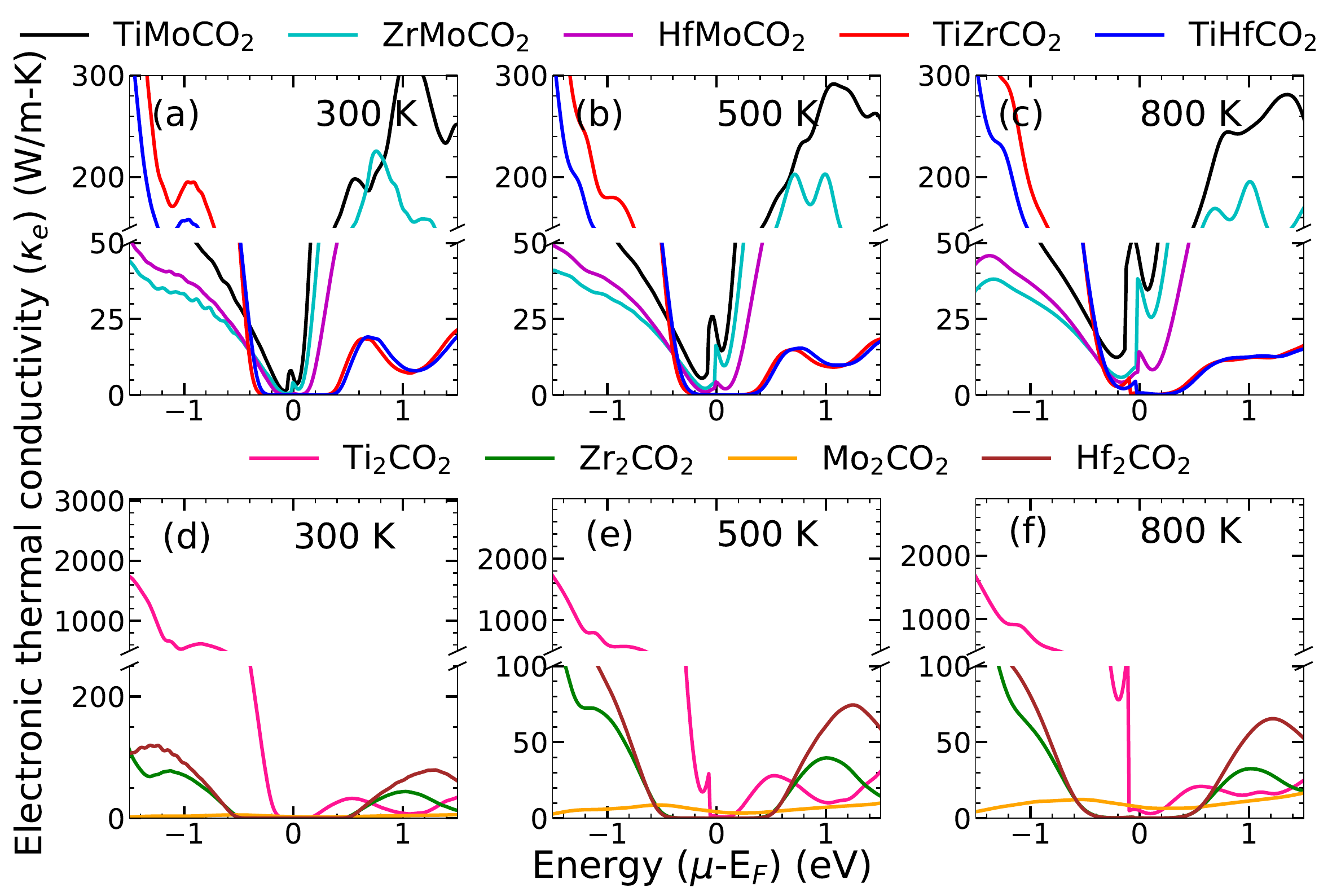}
\caption{Electronic thermal conductivity ($\kappa_{e}$(W/m-K)) as a function of energy at different temperatures. The top(bottom) panel shows results for Janus(Parent) MXenes..}
\label{kappae}
\end{figure}

According to DP theory \cite{bardeen1950deformation}, carrier mobility $\mu$ depends upon the carrier effective mass $m^{*}$, the in-plane stiffness constant $C_{2D}$ and the DP constant $E_{d}$, a measure of electron-phonon coupling strength, the following way: $\mu \propto \frac{C_{2D}}{|m^{*}|^{2} \left(E_{d} \right)^{2}}$. It is expected that $m^{*}$ and $E_{d}$ will have more profound effects on $\mu$. From the calculated values of these quantities presented in Table 2, supplementary material, we infer that the trends in $\mu$ for different charge carriers are dictated by $m^{*}$ which in turn is decided by the curvatures of the top(bottom) of valence(conduction)bands. For example, a noticeably large value of $m^{*}_{e}$ for TiMoCO$_{2}$ in conjunction with small $E_{d}$ is responsible for small $\mu^{e}$. The large value of $m^{*}_{e}$ is due to the flattest bottom of the conduction band among the series of compounds investigated. For TiHfCO$_{2}$ and TiZrCO$_{2}$ Janus compounds, $m^{*}_{e} (m^{*}_{h})$ increase (decrease) in comparison with parent compounds Hf$_{2}$CO$_{2}$ and Zr$_{2}$CO$_{2}$. This behaviour is due to the fact that in Janus compounds bottom of the conduction bands (top of the valence bands) are less(more) dispersive. In case of the three Mo-based Janus, larger values of $m^{*}$, in comparison with Ti$_{2}$CO$_{2}$, Zr$_{2}$CO$_{2}$ and Hf$_{2}$CO$_{2}$,irrespective of charge carrier type, are also an artefact of the band structures near Fermi level.

The electronic transport parameters show two overall trends. First, maximum values of the transport parameters are less upon breaking of inversion symmetry in parent MXenes and second, based upon the quantitative trends, the five Janus compounds can be grouped into two distinct sets: three Mo-based compounds and the remaining two. The trends in the electronic transport parts clearly imply that if only the electronic transport is considered, the inversion symmetry breaking to form Janus MXenes may not lead to higher figure of merit (as compared to parent compounds) as the Seebeck coefficient would be dominant due to its higher power in the expression for $ZT$. Therefore, the contribution of lattice thermal conductivity will be crucial. In the next subsection we explore this in detail.
\subsection{Lattice thermal conductivity}
\begin{figure}
    \centering
    \includegraphics[width=1.0\linewidth]{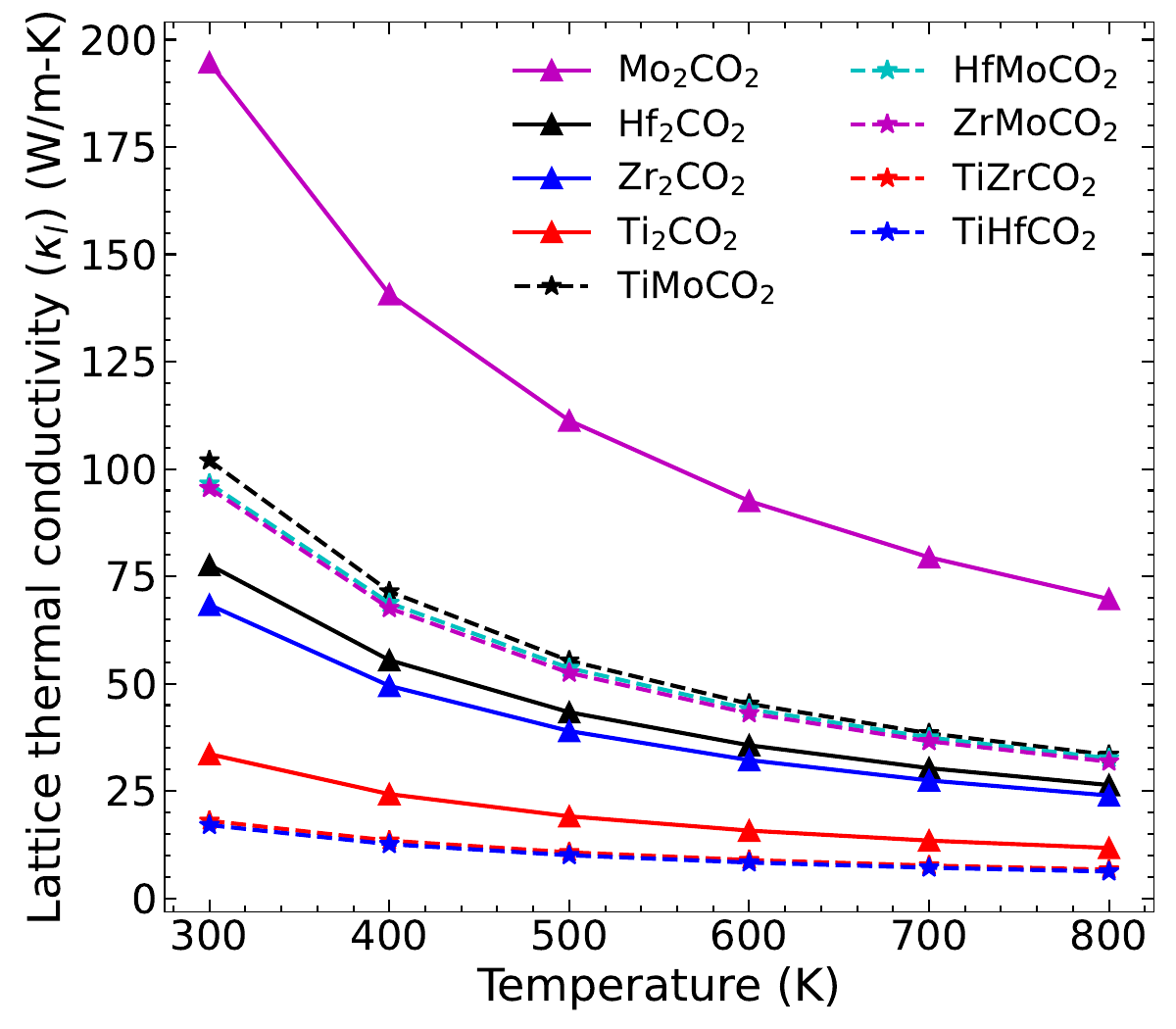}
    \caption{Lattice thermal conductivity ($\kappa_{l}$ (W/m-K)) as function of temperature for the MXenes considered.}
    \label{kappal}
\end{figure}
Figure \ref{kappal} shows the lattice thermal conductivity ($\kappa_{l}$) as a function of  temperature $T$ in the range 300-800 K for the parent and Janus MXenes considered in this work. For all cases $\kappa_{l}$ varies as $(1/T)$, hallmark of intrinsic three-phonon scattering.  Three distinct features are observed in the behaviour of $\kappa_{l}$: (1) Among the parent MXenes, Mo$_{2}$CO$_{2}$ (Ti$_{2}$CO$_{2}$) has the highest (lowest) $\kappa_{l}$ with the following quantitative trend: $\kappa_{l}^{Mo_{2}CO_{2}} >> \kappa_{l}^{Hf_{2}CO_{2}} \gtrsim \kappa_{l}^{Zr_{2}CO_{2}} > \kappa_{l}^{Ti_{2}CO_{2}}$, (2) Among the Janus MXenes, the three Mo-based compounds have higher $\kappa_{l}$ as compared to the remaining two with the quantitative trend as $\kappa_{l}^{TiMoCO_{2}} > \kappa_{l}^{ZrMoCO_{2}} \approx \kappa_{l}^{HfMoCO_{2}} > \kappa_{l}^{TiZrCO_{2}} \gtrsim \kappa_{l}^{TiHfCO_{2}}$ and (3) Thermal conductivities of Mo-free Janus compounds are significantly lower than the relevant parent MXenes ($\kappa_{l}^{TiZrCO_{2}} \approx \kappa_{l}^{TiHfCO_{2}} < \kappa_{l}^{Ti_{2}CO_{2}} << \kappa_{l}^{Zr_{2}CO_{2}} < \kappa_{l}^{Hf_{2}CO_{2}}$) while lattice thermal conductivity of the three Mo-containing Janus are substantially lower than Mo$_{2}$CO$_{2}$ and greater than the other three parent MXenes Zr$_{2}$CO$_{2}$, Hf$_{2}$CO$_{2}$ and Ti$_{2}$CO$_{2}$. The trends are encouraging and intriguing as well. It, therefore, warrants a detailed analysis which has hitherto been unavailable. For example, though lattice thermal conductivities of Ti$_{2}$CO$_{2}$, Zr$_{2}$CO$_{2}$ and Hf$_{2}$CO$_{2}$ have been calculated \cite{gandi2016thermoelectric,sarikurt2018influence} and our results agree well with them, any discussion on the trends based upon quantitative analysis was absent. 

\begin{figure}[H]
    \centering
    \includegraphics[width=1.0\linewidth]{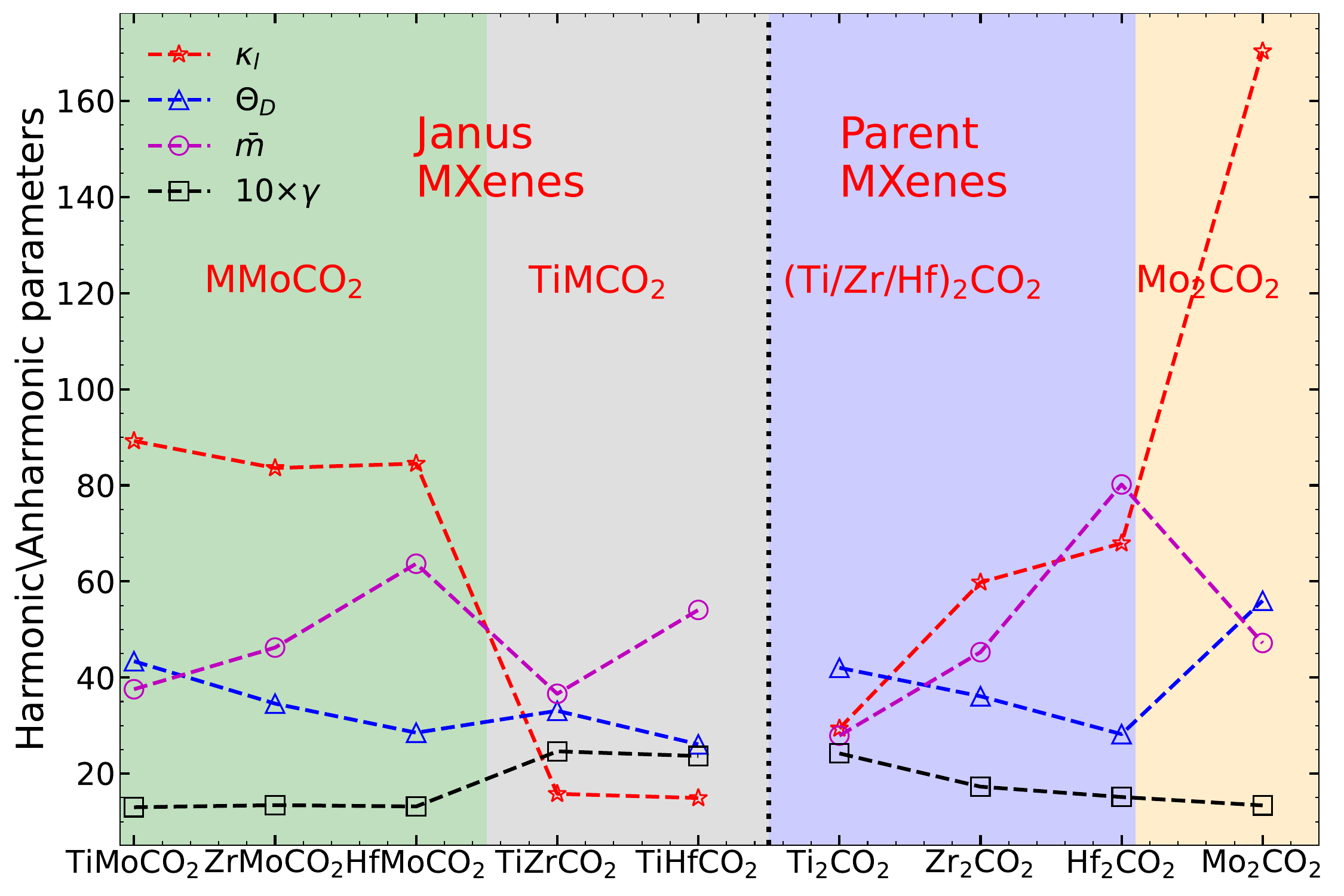}
    \caption{Variation of $\kappa_{l}$ (W/m-K), $\Theta_{D}$ (K), $\bar{m}$ (amu) and $\gamma$ across different MXenes. The results are for 300 K.}
    \label{harmonic}
\end{figure}

Lattice thermal conductivity of non-metallic solids has been understood by a simple qualitative model \cite{slack1973nonmetallic}. According to this model, crystal structure, average atomic mass ($\bar{m}$), bond strength and anharmonicity are the factors deciding $\kappa_{l}$. $\bar{m}$ and the bond strengths that comprise of the harmonic effects influencing $\kappa_{l}$ are manifested through a single quantity, the average acoustic Debye temperature $\Theta_{D}$:  
\begin{equation}
    \frac{1}{\Theta^{3}_{D}} = \frac{1}{3}\Big{(}\frac{1}{\Theta^{3}_{ZA}} + \frac{1}{\Theta^{3}_{TA}} + \frac{1}{\Theta^{3}_{LA}}\Big{)} 
\end{equation}
 $\Theta_{i} = \frac{\hbar \omega^{max}_{i}}{k_{B}}$;  $\omega^{max}_{i}$ is the maximum frequency of $i^{th}$ acoustic mode, $ZA,TA,LA$ are the out-of-plane, transverse and longitudinal acoustic modes, respectively. According to Ref \cite{slack1973nonmetallic}, $\Theta_{D} \propto \kappa_{l}$ and heavier $\bar{m}$ and weak bond strengths will lead to lower value of $\Theta_{D}$, and consequently lower $\kappa_{l}$. Heavier mass and weaker bonding will also lead to lower phonon group velocity and lower $\kappa_{l}$. In Figure \ref{harmonic}, $\bar{m}, \Theta_{D}$ and $\kappa_{l}$ for all MXenes considered in this work are shown. The results depict  few contradictions with the qualitative model \cite{slack1973nonmetallic}. According to the model, Ti$_{2}$CO$_{2}$ should have lattice thermal conductivity larger than Zr$_{2}$CO$_{2}$ and Hf$_{2}$CO$_{2}$. This is because its $\bar{m}$ is the lowest while the bond strengths (Table \ref{tab2}) are not drastically lower in comparison with the other two. The calculated values of $\Theta_{D}$ corroborate this anomaly. In Figure \ref{harmonic} we find that the $\Theta_{D}$ decreases as one moves from Ti$_{2}$CO$_{2}$ to Hf$_{2}$CO$_{2}$ but $\kappa_{l}$ instead of decreasing in the same direction, increases. Similarly, this model cannot explain why $\kappa_{l}^{Mo_{2}CO_{2}}$ is substantially higher than $\kappa_{l}^{Zr_{2}CO_{2}}$ when $\bar{m}$ of the two compounds are almost same. $\Theta_{D}$ of Mo$_{2}$CO$_{2}$ turns out to be substantially higher in comparison with $\Theta_{D}$ of Zr$_{2}$CO$_{2}$ (Figure \ref{harmonic}) providing the explanation for behaviour of $\kappa_{l}$. However, this happens  inspite of no clear trends in the bond strengths. The phonon spectra of these two MXenes (Figure 5, supplementary material) offer some clue to the behaviour of $\Theta_{D}$. The maximum frequencies of the three acoustic modes are noticeably higher in case of Mo$_{2}$CO$_{2}$, leading to larger $\Theta_{D}$ and larger $\kappa_{l}$ as a consequence. Moreover, large gaps between the four-fold degenerate optical mode around 400 cm$^{-1}$ and the two-fold degenerate optical mode around 550 cm$^{-1}$ implies a suppression of phonon-phonon scattering due to decrease in phonon population \cite{ziman2001electrons} leading to subsequent elevation of $\kappa_{l}$. Nevertheless, these are only indirect evidences and do not provide a definite mechanism with quantitative estimates. 

 No such anomaly is encountered if one inspects the two groups of Janus compounds, the Mo-based ones and the remaining two, separately. In both groups, $\bar{m} \propto \frac{1}{\Theta_{D}} \propto \frac{1}{\kappa_{l}}$ is satisfied. The impact of the bond strengths, however, is not clear. The bond strengths of TiZrCO$_{2}$ and TiHfCO$_{2}$ are hardly different so that $\Theta_{D}$ is completely determined by the differences in their masses. The maximum frequencies of the three acoustic modes too explain the behaviour of $\Theta_{D}$. Same is true for the three Mo-based Janus compounds. The results become unexplainable in terms of the simple model with harmonic parameters only when comparison is made between members belonging to different groups. TiMoCO$_{2}$ and TiZrCO$_{2}$, despite having identical $\bar{m}$ have very different $\Theta_{D}$. Like the parent compounds Zr$_{2}$CO$_{2}$ and Mo$_{2}$CO$_{2}$, $\Theta_{D}$ of TiMoCO$_{2}$ is higher, albeit not as substantially as is the case for parent MXenes. This, however, has a profound impact on $\kappa_{l}$; $\kappa_{l}^{TiZrCO_{2}}$ is less than $\kappa_{l}^{TiMoCO_{2}}$ by about 80$\%$ . This trend is found for other compounds too which implies that the quantitative variation cannot be explained in terms of variations in the harmonic parameters alone. 
 
 The limitation of the model appears more pronounced when comparison is made between the lattice thermal conductivities of parent and Janus groups. For example, $\bar{m}^{Zr_{2}CO_{2}}> \bar{m}^{TiZrCO_{2}}> \bar{m}^{Ti_{2}CO_{2}}$ but $\Theta_{D}^{Ti_{2}CO_{2}} > \Theta_{D}^{TiZrCO_{2}} < \Theta_{D}^{Zr_{2}CO_{2}}$. Though the trend in $\Theta_{D}$ is consistent with the trend in $\kappa_{l}$ for these three compounds, the trends in $\bar{m}$ and $\Theta_{D}$ contradict the harmonic model. Even the trends in the bond strengths (Table \ref{tab2}) contradict the model. A comparison between Mo$_{2}$CO$_{2}$ and Mo-based Janus compounds too corroborate this. For this group of MXenes, $\Theta_{D}$ and $\kappa_{l}$ of Mo$_{2}$CO$_{2}$ are always greater than the Janus compounds inspite of the $\bar{m}$ of Janus compounds being equal to (in case of ZrMoCO$_{2}$) or greater (in case of HfMoCO$_{2}$) than $\bar{m}$ of Mo$_{2}$CO$_{2}$. The Janus bonds too are stronger. 
 
 One more hindrance in understanding the trends of harmonic parameters and $\kappa_{l}$ across all parents and Janus compounds using the harmonic model \cite{slack1973nonmetallic} is that the symmetry of the members of two different groups of compounds (one containing Mo and the other consists of the remaining ones) is different due to different site preference of the -O functional group. Therefore, we next compute the phonon group velocities as they are directly proportional to lattice thermal conductivity (Equation (\ref{kl_eq})). The results are presented in Figures S7 and S8, supplementary material. The results still cannot resolve the anomalies. For example, Hf$_{2}$CO$_{2}$ has group velocities lower than Ti$_{2}$CO$_{2}$ and  TiHfCO$_{2}$, HfMoCO$_{2}$ have group velocities comparable to TiZrCO$_{2}$. But the trends in their $\kappa_{l}$ do not support this behaviour. 
 
 These anomalies, therefore, indicate that anharmonicity plays a key role in understanding the trends in $\kappa_{l}$. Earlier works on parent compounds Ti$_{2}$CO$_{2}$, Zr$_{2}$CO$_{2}$ and Hf$_{2}$CO$_{2}$ \cite{gandi2016thermoelectric,sarikurt2018influence} indicated this by analysing the phonon dispersions. However, they did not analyse the reasons behind the trend among the compounds , neither did they quantify the extent of anharmonicity.  In what follows, we compute various quantities associated with the anharmonicity in the systems. In Figure \ref{harmonic}, we show the total Gruneisen parameter $\gamma$ , the measure of the strength of anharmonicity in the system, calculated using phonon BTE \cite{li2014shengbte}. Among the parent MXenes, $\gamma$ decreases monotonically from Ti$_{2}$CO$_{2}$ to Mo$_{2}$CO$_{2}$ suggesting that largest (smallest) anharmonicity in Ti$_{2}$CO$_{2}$ (Mo$_{2}$CO$_{2}$) can be the reason for the smallest(largest) $\kappa_{l}$ in Ti$_{2}$CO$_{2}$(Mo$_{2}$CO$_{2}$). The comparable $\gamma$ values of Zr$_{2}$CO$_{2}$ and Hf$_{2}$CO$_{2}$ explains their near identical $\kappa_{l}$ that are between Mo$_{2}$CO$_{2}$ and Ti$_{2}$CO$_{2}$. Among the Janus MXenes, larger anharmonicity is found in compounds without Mo, with their $\gamma$ values being largest among all parent and Janus considered. $\gamma$ for three Mo-based MXenes are almost identical that fits perfectly into the trends of their $\kappa_{l}$. A comparison between $\gamma$ of parent and Janus MXenes too perfectly correlate with the trends in the lattice thermal conductivities.  

\begin{figure}[H]
    \centering
    \includegraphics[width=1.0\linewidth]{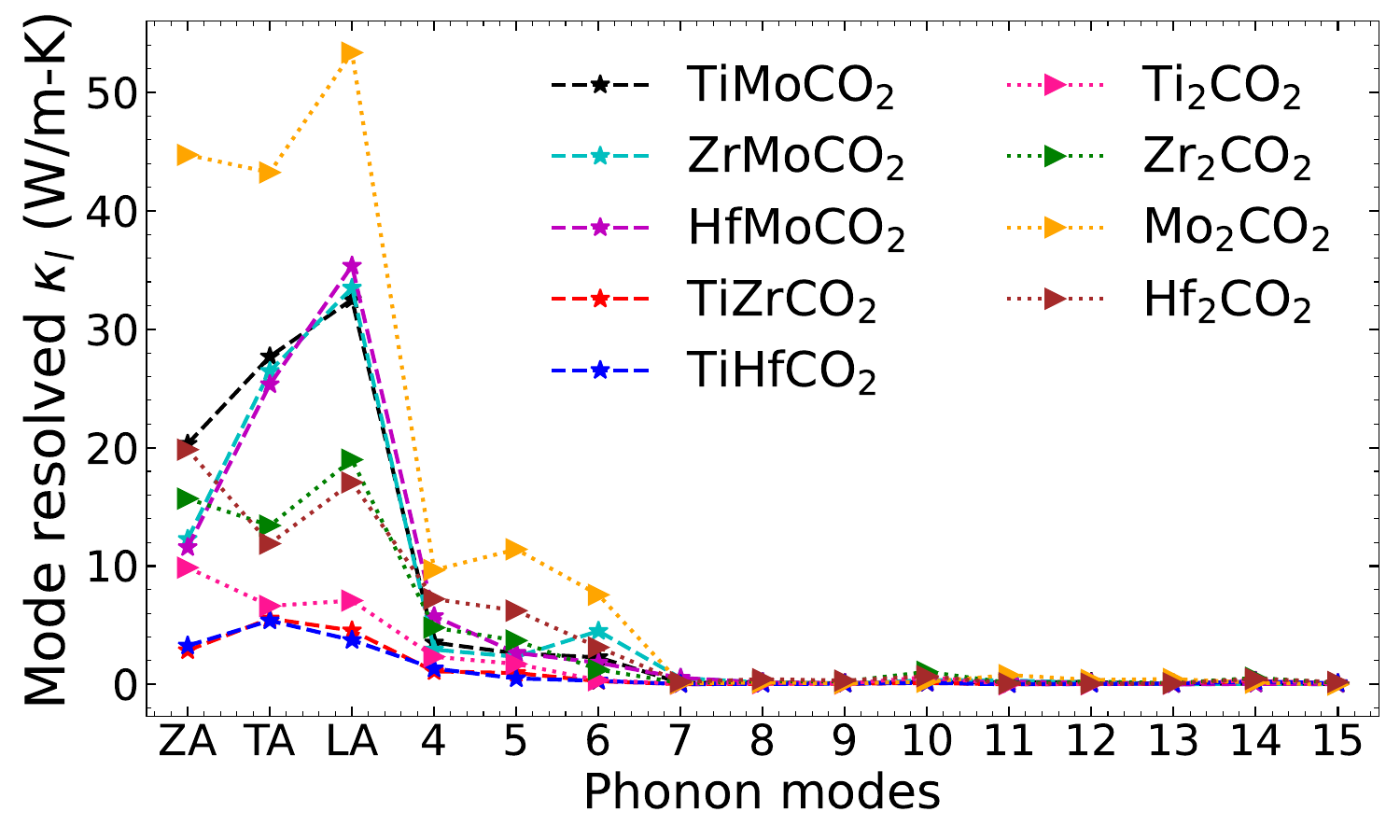}
    \caption{Mode resolved lattice thermal conductivity for all  MXenes considered. The 3 acoustic and 12 optical modes in ascending order of frequency are plotted along $x$-axis.}
    \label{mode}
\end{figure}

In order to completely understand the extent of anharmonic effects and develop a microscopic picture, we next look into the phonon modes responsible and the strengths of phonon phonon scattering. To this end, we first look at the frequency range and the corresponding modes that are major contributors to $\kappa_{l}$. In Figure 9, supplementary material, we show the variations in normalised cumulative lattice thermal conductivity ($\kappa_{c}$/$\kappa_{l}$) as a function of frequency. The results are presented for 300K only. The results show that for all systems considered, phonon modes with frequencies upto 200 cm$^{-1}$ are the major contributors to $\kappa_{l}$. The mode resolved $\kappa_{l}$, presented in Figure \ref{mode}, corroborates this. We find that for all systems the three acoustic and first three optical modes (numbered 4,5,6 in the figure) contribute almost entirely to the lattice thermal conductivity. A noticeable feature is that for Mo$_{2}$CO$_{2}$, the contributions from acoustic and optical modes towards $\kappa_{l}$ are significantly different. This is also true for the Janus compounds containing Mo. On the other hand, Ti$_{2}$CO$_{2}$, TiZrCO$_{2}$ and TiHfCO$_{2}$ have near equal contributions from acoustic and optical modes implying strong phonon-phonon interactions.
To connect the behaviour of mode-resolved $\kappa_{l}$ with anharmonicity, we first look at mode-resolved Gr\"uneisen parameters ($\gamma_{\lambda}$) calculated using third-order IFCs \cite{gruneisen}. The results are presented in Figure 10, supplementary material. We find that among the parent compounds, Zr$_{2}$CO$_{2}$ has  larger $\gamma_{\lambda}$ than Mo$_{2}$CO$_{2}$ in the relevant frequency range, implying significant large anharmonicity in the former. This can be correlated to their very different $\kappa_{l}$ with $\kappa_{l}^{Zr_{2}CO_{2}}$ much smaller than that of the other. $\gamma_{\lambda}$ for Zr$_{2}$CO$_{2}$ and Hf$_{2}$CO$_{2}$ in the same frequency range are comparable. Ti$_{2}$CO$_{2}$ has the highest $\gamma_{\lambda}$ among all parents. The $\gamma_{\lambda}$ in non-Mo Janus compounds are 2-3 times higher than Ti$_{2}$CO$_{2}$. The Mo-based Janus compounds, on the other hand, have extremely low $\gamma_{\lambda}$. These trends nicely explain the behaviour of $\kappa_{l}$. In fact $\gamma_{\lambda}$ turns out to be a better indicator than $\gamma$ for comparing the degree of anharmonicity among compounds. For example, $\gamma$ of Ti$_{2}$CO$_{2}$ and TiHfCO$_{2}$ are same. But $\gamma_{\lambda}$ of the later is much higher in the relevant frequency range. The differences in their $\kappa_{l}$ values, thus, can only be explained by $\gamma_{\lambda}$. 

\begin{figure}
	\centering
	\includegraphics[width=1.0\linewidth]{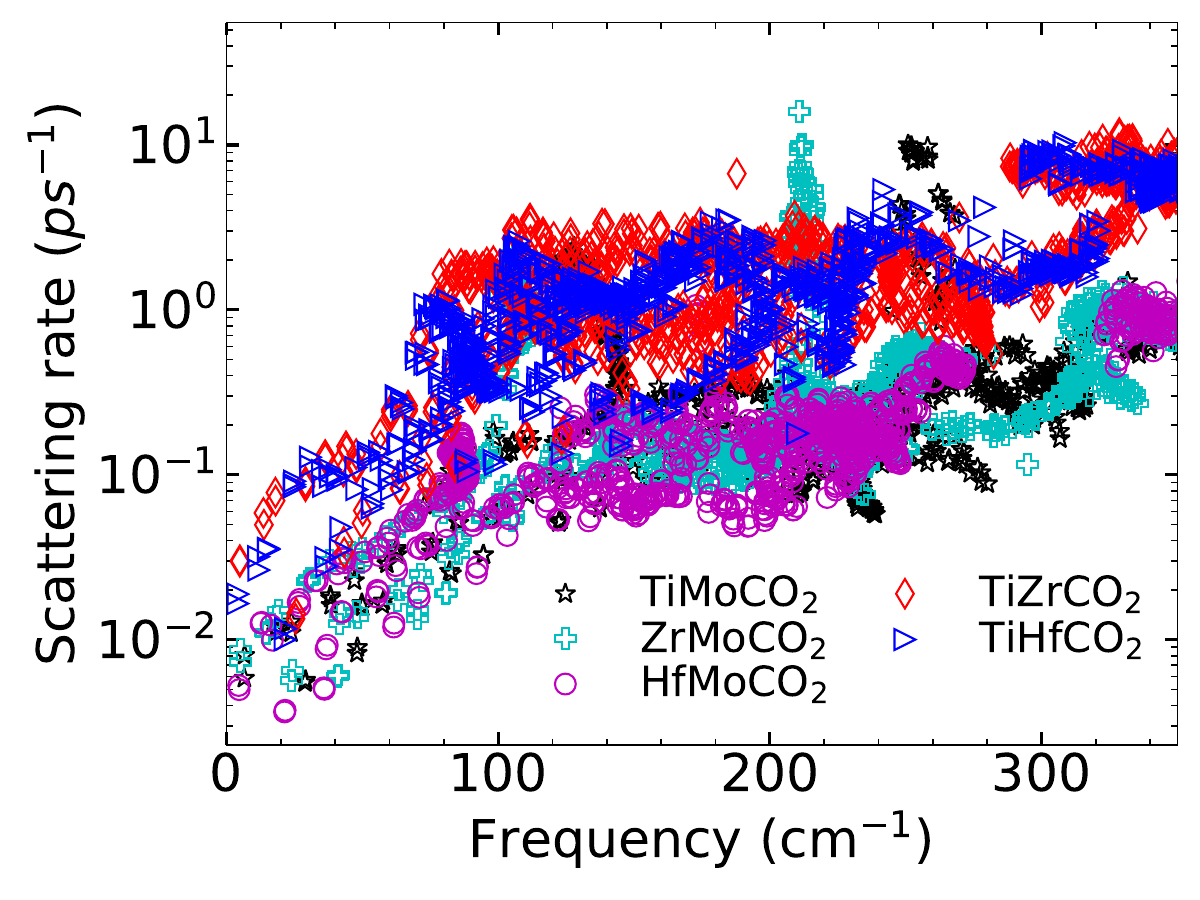}
	\caption{Anharmonic scattering rates as  function of frequency for Janus MXenes considered .}
	\label{scat_js}
\end{figure}

\begin{figure}
	\centering
	\includegraphics[width=1.0\linewidth]{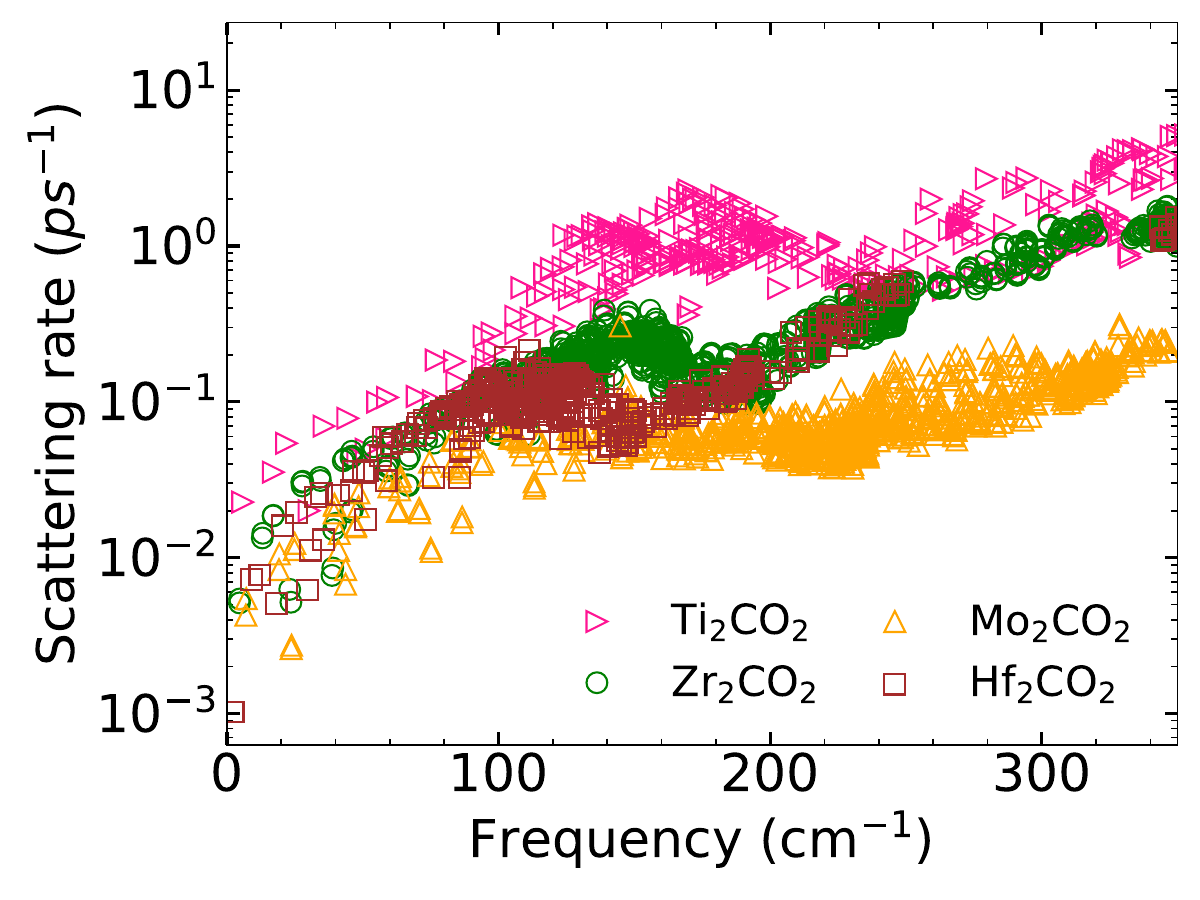}
	\caption{Anharmonic scattering rates as function of phonon frequency for parent MXenes considered.}
	\label{scat_ps}
\end{figure}

We further understand the degree of anharmonic effects as quantified by $\gamma_{\lambda}$ by calculating the anharmonic scattering rates over the frequency region. The results for Janus (parent) MXenes are shown in Figure \ref{scat_js}(Figure \ref{scat_ps}). Among the parent compounds, Ti$_{2}$CO$_{2}$ has appreciably large scattering rates in comparison with other three. Mo$_{2}$CO$_{2}$ has the lowest scattering rate, scattering rates of the other two lie in between these two. The large scattering rate of Ti$_{2}$CO$_{2}$ finally resolves the anomaly regarding its lowest $\kappa_{l}$ as compared to other parent compounds despite all harmonic parameters indicating the opposite. The lattice thermal conductivity is directly proportional to the group velocity $v_{g}$ and the phonon relaxation time $\tau^{0}$ (Equation \ref{kl_eq}). Since scattering rate is inversely proportional to $\tau^{0}$, the effect of larger $v_{g}$ of Ti$_{2}$CO$_{2}$ discussed earlier is nullified by the small $\tau^{0}$, yielding lowest $\kappa_{l}$ among the parent compounds. Among Janus compounds, higher scattering rates are observed for TiZrCO$_{2}$ and TiHfCO$_{2}$. This explains their higher values of $\gamma_{\lambda}$  and consequently lower $\kappa_{l}$ among the MXenes considered. In Figure 11, supplementary material, comparison between scattering rates in Janus and parent compounds are made. The noticeable trend observed in this comparison is that while the scattering rates of MMoCO$_{2}$ Janus compounds lie in between that of M$_{2}$CO$_{2}$ and Mo$_{2}$CO$_{2}$, the scattering rates of TiZrCO$_{2}$ and TiHfCO$_{2}$ are higher than the corresponding parent MXenes. To find the reason, we first look at the accessible phase space for phonon scattering since $\tau^{0}$ ( and thus $\kappa_{l}$) is inversely proportional to the accessible phase space \cite{lindsay2008three,pandey2017lattice}which is determined by the condition of phonon energy and momentum conservation.    
\begin{figure}
    \centering
    \includegraphics[width=1.0\linewidth]{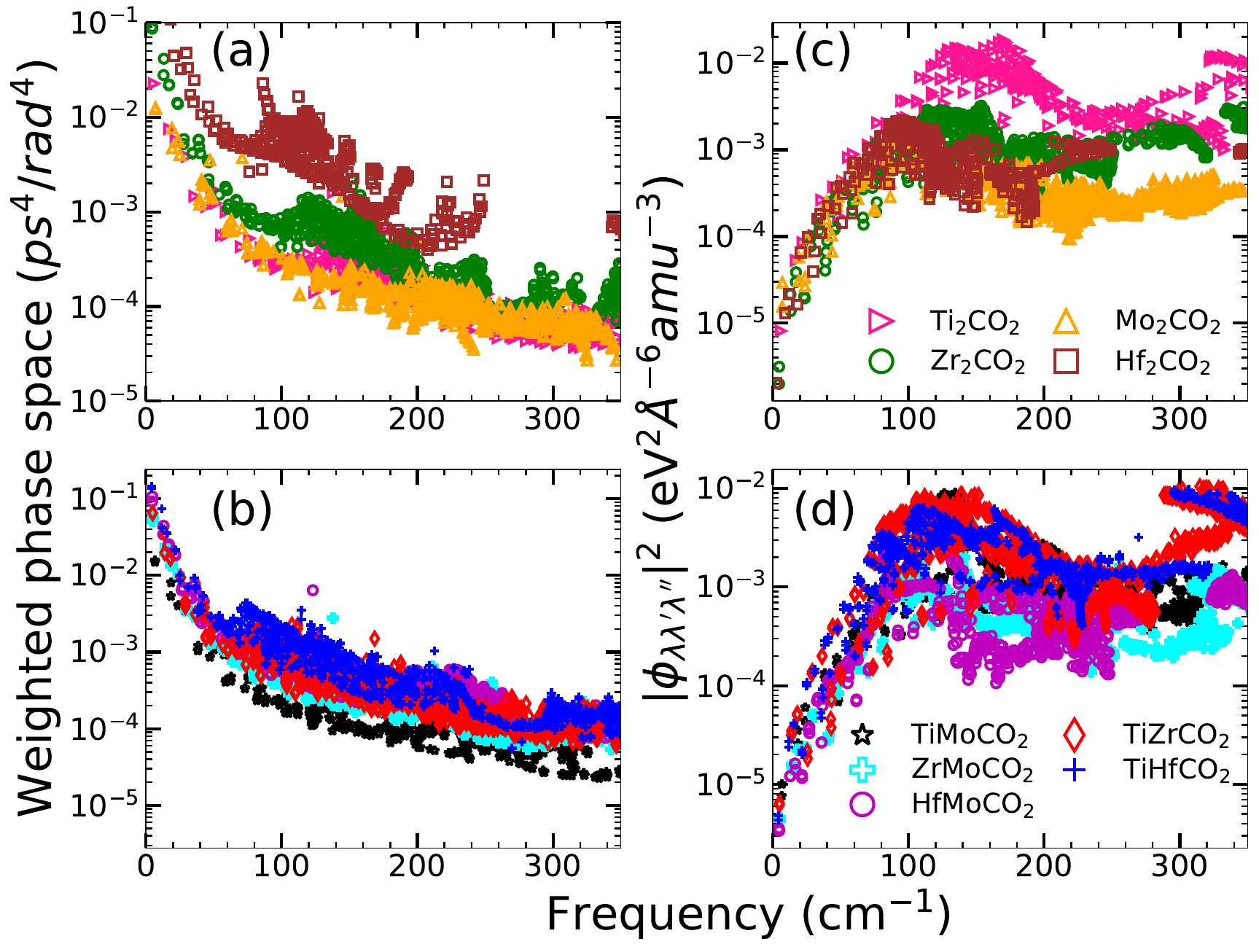}
    \caption{(a,b) Available weighted phase space as  function of phonon frequency for the MXenes considered. (c,d) Average scattering matrix element $|\phi_{\lambda\lambda^{'}\lambda^{"}}^{\pm}|^{2}$ as  function of frequency for all  MXenes considered}
    \label{wp}
\end{figure}

Figure \ref{wp}(a)((b)) shows the weighted scattering phase space (possible three-phonon scattering processes weighted by the frequencies) accessible to the  parent(Janus) MXenes.  Among the Janus MXenes, available phase spaces for Mo-based ones are lower as compared to the others. TiHfCO$_{2}$ has more accessible phase space than TiZrCO$_{2}$. More phase spaces available for three-phonon scattering, thus, can be the explanation for higher scattering rates seen in these two Janus MXenes. But, this proposition is violated in case of the parent MXenes. We find that Ti$_{2}$CO$_{2}$ has less accessible phase space compared to Hf$_{2}$CO$_{2}$ and Zr$_{2}$CO$_{2}$. Even TiHfCO$_{2}$ has less accessible phase space in comparison with Hf$_{2}$CO$_{2}$ in the frequency range 100-150 cm$^{-1}$, the range most responsible for anharmonic scattering. Therefore, the trends in the anharmonic scattering rates cannot be understood in terms of accessible phase space. To resolve this issue, we present the results on the strength of scattering matrix elements $|\phi_{\lambda\lambda^{'}\lambda^{"}}^{\pm}|^{2}$, which is a measure of the strength of three phonon scattering and depends on the anharmonic IFCs (Equation (\ref{eq5})). Figure \ref{wp}(c)(Figure \ref{wp}(d)) show the average scattering matrix elements as a function of frequency for parent(Janus) MXenes. It is clear from Fig.\ref{wp}(c) that throughout the frequency range, the scattering strength in Ti$_{2}$CO$_{2}$ is more than Hf$_{2}$CO$_{2}$. Therefore, despite Hf$_{2}$CO$_{2}$ having more available phase space than Ti$_{2}$CO$_{2}$, it has lower scattering rate. Among the Janus compounds, TiZrCO$_{2}$ has the largest scattering strength with TiHfCO$_{2}$ competing closely. Therefore, the degree of anharmonicity among the parent and among the Janus can be finally understood in terms of the anharmonic scattering strengths. Comparison between parents and Janus, however, shows an anomaly : scattering strengths in Ti$_{2}$CO$_{2}$ in the frequency range of interest is still slightly higher than those in TiZrCO$_{2}$. Despite this, larger anharmonicity in TiZrCO$_{2}$ can be attributed to higher number of scatterings as can be understood by comparing Figures \ref{wp} (c) and (d). 
   
The inversion symmetry breaking in the Janus MXenes produces dispersions in the bond strengths (Table \ref{tab2}). We find that such dispersion can be connected to the degree of anharmonicity in Janus. The phonon densities of states of TiHfCO$_{2}$(Figure \ref{phonon}(e)) shows that in the frequency range 100-200 cm$^{-1}$, the vibrations are dominated by the Ti atoms. In this Janus Ti-anion bond strengths weaken considerably as compared to the corresponding ones in parent Ti$_{2}$CO$_{2}$. The Hf-cation bond strengths strengthen, on the other hand, in comparison to those in parent MXene Hf$_{2}$CO$_{2}$. In Janus TiZrCO$_{2}$ ( Figure \ref{phonon}(d)), both Ti and Zr vibrations contribute in the relevant frequency range. Here too we find significant dispersion in bond strengths, similar to TiHfCO$_{2}$. In the Mo-based Janus ZrMoCO$_{2}$ and HfMoCO$_{2}$, these dispersions are much less. Although dispersions are there in TiMoCO$_{2}$, the Ti-cation bond strengths have hardly changed in comparison with Ti$_{2}$CO$_{2}$. Since the vibrations in the frequency range of interest is overwhelmingly dominated by the Ti atoms, the Ti-anion bond strengths decide the extent of anharmonicity.   

\subsection{Figure of Merit}
\begin{figure}
    \centering
    \includegraphics[width=1.0\linewidth]{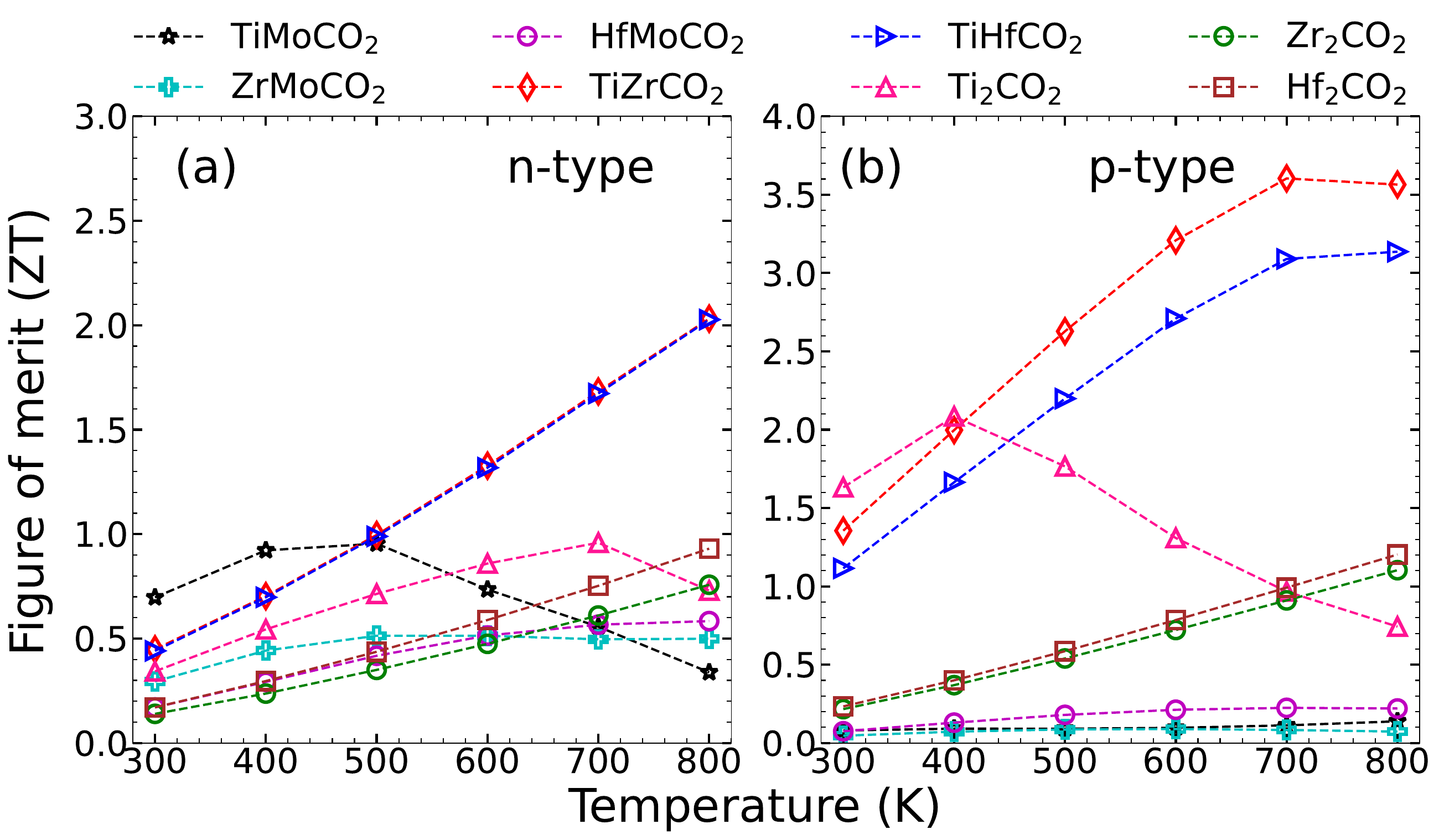}
    \caption{Maximum Figure of merit ($ZT$) of all MXenes considered, as function of temperature. Results for both  n and p-doped systems are shown.}
    \label{zt}
\end{figure}

Figure \ref{zt} shows the maximum value of  figure of Merit $ZT$ as a function of temperature for all parent and Janus MXenes. Energy window from -1.5 to 1.5 eV and carrier concentration $\sim$ 10$^{20}$ cm$^{-3}$ are used to extract maximum $ZT$. We find that surface engineering by forming Janus greatly enhances the $ZT$. The maximum $ZT$ of 3.5 at 800K  is obtained for TiZrCO$_{2}$ in case of p-doped systems. This is more than double the maximum $ZT$ obtained in any of the parent MXenes (a maximum $ZT$ value of 1.6 at 300 K is obtained for Ti$_{2}$CO$_{2}$). p-doped TiHfCO$_{2}$ also yields a high $ZT$ of $\sim 3$ at 800K. These two Janus compounds, even when $n$-doped, produce maximum $ZT$ of $2$ while none of the parent compounds and Mo-based Janus could produce a maximum $ZT$ more than $1$. Dominant reason of such high $ZT$ in these two systems, irrespective of type of doping, is due to their extremely low $\kappa_{l}$. High values of $S$ and $\sigma$ for p-doped TiZrCO$_{2}$ and TiHfCO$_{2}$ are responsible for substantially higher $ZT$ in these systems when they are doped by p-type carriers in comparison to doping by n-type one. Thus, desired values of maximum $ZT$ can be obtained by breaking the inversion symmetry in MXenes.    
\section{Thermodynamical stability}
\begin{figure}
    \centering
    \includegraphics[width=1.0\linewidth]{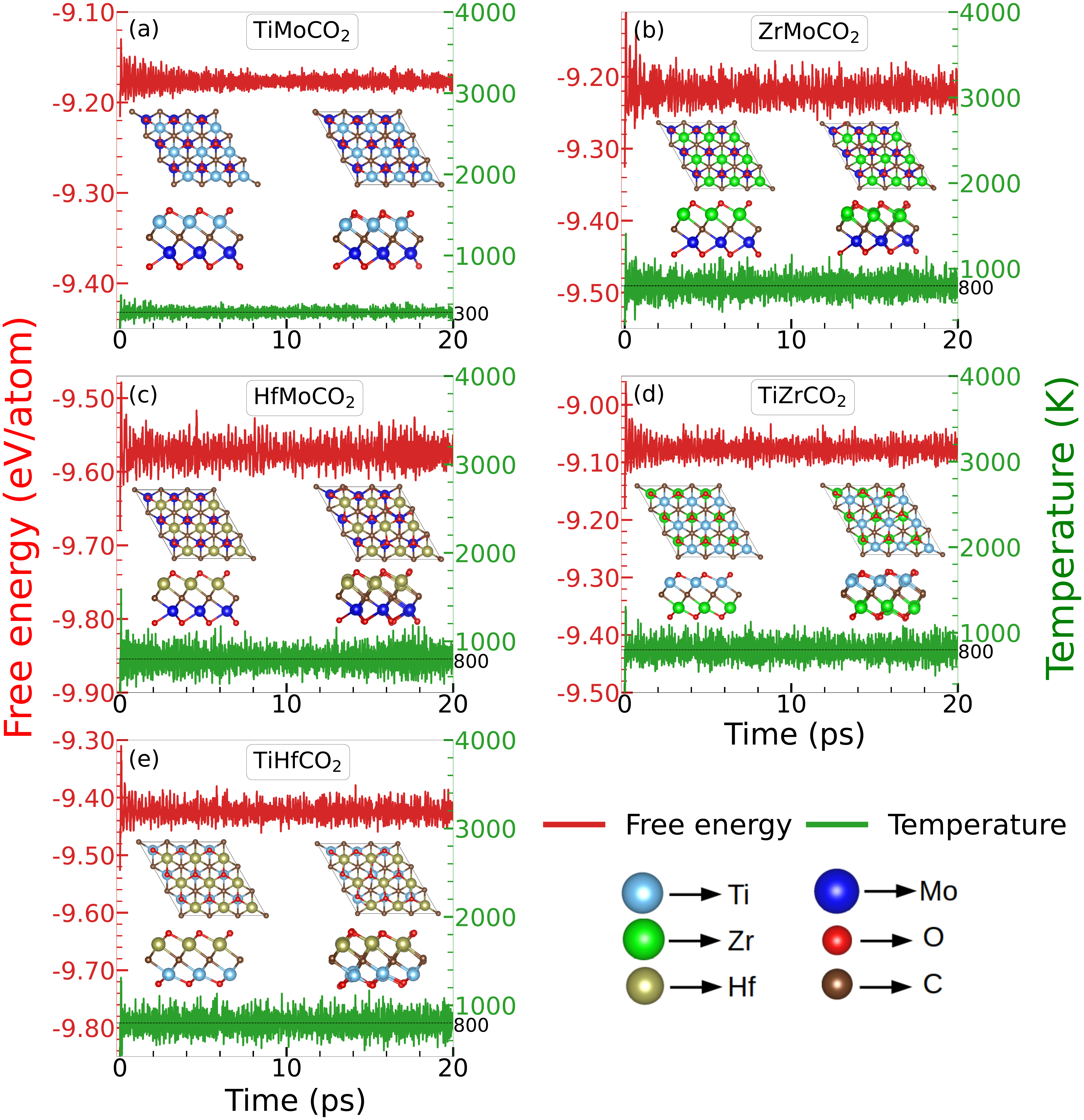}
    \caption{AIMD simulations for the Janus MXenes considered.}
    \label{aimd}
\end{figure}
Since the Janus compounds presented in this work are yet to be synthesised experimentally, it is important to check their thermodynamic stabilities at the maximum temperature at which these are proposed to be used as thermoelectric devices. To this end, we have performed the {\it ab initio} molecular dynamics (AIMD) simulations for all five Janus MXene considered. The simulations are performed at 800K, the maximum temperature at which thermoelectric parameters are calculated. Figure \ref{aimd} shows the fluctuations in Free energy (red curves ) and temperature (green curves). The top and side views of the initial (at T$=$0 K) and final structures (at T$=$800 K) are shown in the insets. The Free energies and the temperatures fluctuate about their average values and remain constant for the simulation time cycle of 20 ps. No signature of bond distortion or loss of structural symmetry are observed, which ensures that these Janus MXenes possess good thermal stability and can be used for thermoelectric applications at significantly high temperatures.
\section{Conclusions}
Inversion symmetry breaking in two-dimensional materials has turned out to be the driving force behind many a material properties. In the context of thermoelectric properties of materials, this aspect has not been addressed in detail. In this work, the interplay of symmetry breaking, electronic structure, lattice dynamics and transport properties has been investigated with the help of DFT based first-principles techniques and semi-classical Boltzmann transport theory. The systems chosen are from MXene family whose compositional flexibility is ideal to study such structure-property relationships. The inversion symmetry in M$_{2}$CO$_{2}$ MXenes is broken by manipulating its surfaces through substitution of transition metal atoms resulting in Janus MXenes. Our calculations predict two Janus compounds, TiZrCO$_{2}$ and TiHfCO$_{2}$ with thermoelectric figure of merit $\sim 3$, much larger in comparison with M$_{2}$CO$_{2}$ MXenes considered. In the MXene family of compounds, only two have been reported to possess such large figure of merit. With great detail, we have systematically done a comparative study of the thermoelectric parameters of the Janus and corresponding parent MXenes to develop a microscopic understanding of the trends in the thermoelectric parameters. We infer that the dispersions in the bond strengths, and weakening of metal-anion bonds on one particular surface, due to lowering of symmetry, are responsible for the degree of anharmonicity in the Janus compounds. This, in turn, affects their lattice thermal conductivities in particular and the thermoelectric figure of merit subsequently. The importance of this work lies in the fact that it opens up possibilities to exploit the tunability of the MXene compounds by manipulating their surfaces and improve their thermoelectric performances manifold.     
    
\begin{acknowledgments}
The authors gratefully acknowledge the Department of Science and Technology, India, for the computational facilities under Grant No. SR/FST/P-II/020/2009 and IIT Guwahati
for the PARAM supercomputing facility.
\end{acknowledgments}


\begin{thebibliography}{51}%
\makeatletter
\providecommand \@ifxundefined [1]{%
 \@ifx{#1\undefined}
}%
\providecommand \@ifnum [1]{%
 \ifnum #1\expandafter \@firstoftwo
 \else \expandafter \@secondoftwo
 \fi
}%
\providecommand \@ifx [1]{%
 \ifx #1\expandafter \@firstoftwo
 \else \expandafter \@secondoftwo
 \fi
}%
\providecommand \natexlab [1]{#1}%
\providecommand \enquote  [1]{``#1''}%
\providecommand \bibnamefont  [1]{#1}%
\providecommand \bibfnamefont [1]{#1}%
\providecommand \citenamefont [1]{#1}%
\providecommand \href@noop [0]{\@secondoftwo}%
\providecommand \href [0]{\begingroup \@sanitize@url \@href}%
\providecommand \@href[1]{\@@startlink{#1}\@@href}%
\providecommand \@@href[1]{\endgroup#1\@@endlink}%
\providecommand \@sanitize@url [0]{\catcode `\\12\catcode `\$12\catcode
  `\&12\catcode `\#12\catcode `\^12\catcode `\_12\catcode `\%12\relax}%
\providecommand \@@startlink[1]{}%
\providecommand \@@endlink[0]{}%
\providecommand \url  [0]{\begingroup\@sanitize@url \@url }%
\providecommand \@url [1]{\endgroup\@href {#1}{\urlprefix }}%
\providecommand \urlprefix  [0]{URL }%
\providecommand \Eprint [0]{\href }%
\providecommand \doibase [0]{http://dx.doi.org/}%
\providecommand \selectlanguage [0]{\@gobble}%
\providecommand \bibinfo  [0]{\@secondoftwo}%
\providecommand \bibfield  [0]{\@secondoftwo}%
\providecommand \translation [1]{[#1]}%
\providecommand \BibitemOpen [0]{}%
\providecommand \bibitemStop [0]{}%
\providecommand \bibitemNoStop [0]{.\EOS\space}%
\providecommand \EOS [0]{\spacefactor3000\relax}%
\providecommand \BibitemShut  [1]{\csname bibitem#1\endcsname}%
\let\auto@bib@innerbib\@empty
\bibitem [{\citenamefont {Sootsman}\ \emph {et~al.}(2009)\citenamefont
  {Sootsman}, \citenamefont {Chung},\ and\ \citenamefont
  {Kanatzidis}}]{sootsman2009new}%
  \BibitemOpen
  \bibfield  {author} {\bibinfo {author} {\bibfnamefont {J.}~\bibnamefont
  {Sootsman}}, \bibinfo {author} {\bibfnamefont {D.}~\bibnamefont {Chung}}, \
  and\ \bibinfo {author} {\bibfnamefont {M.}~\bibnamefont {Kanatzidis}},\
  }\href {\doibase https://doi.org/10.1002/anie.200900598} {\bibfield
  {journal} {\bibinfo  {journal} {Angew. Chem. Int. Ed.}\ }\textbf {\bibinfo
  {volume} {48}},\ \bibinfo {pages} {8616} (\bibinfo {year}
  {2009})}\BibitemShut {NoStop}%
\bibitem [{\citenamefont {DiSalvo}(1999)}]{disalvo1999thermoelectric}%
  \BibitemOpen
  \bibfield  {author} {\bibinfo {author} {\bibfnamefont {F.~J.}\ \bibnamefont
  {DiSalvo}},\ }\href {\doibase 10.1126/science.285.5428.703} {\bibfield
  {journal} {\bibinfo  {journal} {Science}\ }\textbf {\bibinfo {volume}
  {285}},\ \bibinfo {pages} {703} (\bibinfo {year} {1999})}\BibitemShut
  {NoStop}%
\bibitem [{\citenamefont {Yu}\ \emph {et~al.}(2019)\citenamefont {Yu},
  \citenamefont {Li}, \citenamefont {Nie}, \citenamefont {Zhang},\ and\
  \citenamefont {Sun}}]{yu2019ultralow}%
  \BibitemOpen
  \bibfield  {author} {\bibinfo {author} {\bibfnamefont {J.}~\bibnamefont
  {Yu}}, \bibinfo {author} {\bibfnamefont {T.}~\bibnamefont {Li}}, \bibinfo
  {author} {\bibfnamefont {G.}~\bibnamefont {Nie}}, \bibinfo {author}
  {\bibfnamefont {B.-P.}\ \bibnamefont {Zhang}}, \ and\ \bibinfo {author}
  {\bibfnamefont {Q.}~\bibnamefont {Sun}},\ }\href {\doibase
  10.1039/C9NR01501A} {\bibfield  {journal} {\bibinfo  {journal} {Nanoscale}\
  }\textbf {\bibinfo {volume} {11}},\ \bibinfo {pages} {10306} (\bibinfo {year}
  {2019})}\BibitemShut {NoStop}%
\bibitem [{\citenamefont {Biswas}\ \emph {et~al.}(2012)\citenamefont {Biswas},
  \citenamefont {He}, \citenamefont {Blum}, \citenamefont {Wu}, \citenamefont
  {Hogan}, \citenamefont {Seidman}, \citenamefont {Dravid},\ and\ \citenamefont
  {Kanatzidis}}]{biswas2012high}%
  \BibitemOpen
  \bibfield  {author} {\bibinfo {author} {\bibfnamefont {K.}~\bibnamefont
  {Biswas}}, \bibinfo {author} {\bibfnamefont {J.}~\bibnamefont {He}}, \bibinfo
  {author} {\bibfnamefont {I.~D.}\ \bibnamefont {Blum}}, \bibinfo {author}
  {\bibfnamefont {C.-I.}\ \bibnamefont {Wu}}, \bibinfo {author} {\bibfnamefont
  {T.~P.}\ \bibnamefont {Hogan}}, \bibinfo {author} {\bibfnamefont {D.~N.}\
  \bibnamefont {Seidman}}, \bibinfo {author} {\bibfnamefont {V.~P.}\
  \bibnamefont {Dravid}}, \ and\ \bibinfo {author} {\bibfnamefont {M.~G.}\
  \bibnamefont {Kanatzidis}},\ }\href {\doibase 10.1038/nature11439} {\bibfield
   {journal} {\bibinfo  {journal} {Nature}\ }\textbf {\bibinfo {volume}
  {489}},\ \bibinfo {pages} {414} (\bibinfo {year} {2012})}\BibitemShut
  {NoStop}%
\bibitem [{\citenamefont {Hicks}\ \emph {et~al.}(1996)\citenamefont {Hicks},
  \citenamefont {Harman}, \citenamefont {Sun},\ and\ \citenamefont
  {Dresselhaus}}]{hicks1996experimental}%
  \BibitemOpen
  \bibfield  {author} {\bibinfo {author} {\bibfnamefont {L.~D.}\ \bibnamefont
  {Hicks}}, \bibinfo {author} {\bibfnamefont {T.~C.}\ \bibnamefont {Harman}},
  \bibinfo {author} {\bibfnamefont {X.}~\bibnamefont {Sun}}, \ and\ \bibinfo
  {author} {\bibfnamefont {M.~S.}\ \bibnamefont {Dresselhaus}},\ }\href
  {\doibase 10.1103/PhysRevB.53.R10493} {\bibfield  {journal} {\bibinfo
  {journal} {Phys. Rev. B}\ }\textbf {\bibinfo {volume} {53}},\ \bibinfo
  {pages} {R10493} (\bibinfo {year} {1996})}\BibitemShut {NoStop}%
\bibitem [{\citenamefont {Hicks}\ and\ \citenamefont
  {Dresselhaus}(1993)}]{hicks1993effect}%
  \BibitemOpen
  \bibfield  {author} {\bibinfo {author} {\bibfnamefont {L.~D.}\ \bibnamefont
  {Hicks}}\ and\ \bibinfo {author} {\bibfnamefont {M.~S.}\ \bibnamefont
  {Dresselhaus}},\ }\href {\doibase 10.1103/PhysRevB.47.12727} {\bibfield
  {journal} {\bibinfo  {journal} {Phys. Rev. B}\ }\textbf {\bibinfo {volume}
  {47}},\ \bibinfo {pages} {12727} (\bibinfo {year} {1993})}\BibitemShut
  {NoStop}%
\bibitem [{\citenamefont {Dresselhaus}\ \emph {et~al.}(2007)\citenamefont
  {Dresselhaus}, \citenamefont {Chen}, \citenamefont {Tang}, \citenamefont
  {Yang}, \citenamefont {Lee}, \citenamefont {Wang}, \citenamefont {Ren},
  \citenamefont {Fleurial},\ and\ \citenamefont {Gogna}}]{dresselhaus2007new}%
  \BibitemOpen
  \bibfield  {author} {\bibinfo {author} {\bibfnamefont {M.}~\bibnamefont
  {Dresselhaus}}, \bibinfo {author} {\bibfnamefont {G.}~\bibnamefont {Chen}},
  \bibinfo {author} {\bibfnamefont {M.}~\bibnamefont {Tang}}, \bibinfo {author}
  {\bibfnamefont {R.}~\bibnamefont {Yang}}, \bibinfo {author} {\bibfnamefont
  {H.}~\bibnamefont {Lee}}, \bibinfo {author} {\bibfnamefont {D.}~\bibnamefont
  {Wang}}, \bibinfo {author} {\bibfnamefont {Z.}~\bibnamefont {Ren}}, \bibinfo
  {author} {\bibfnamefont {J.-P.}\ \bibnamefont {Fleurial}}, \ and\ \bibinfo
  {author} {\bibfnamefont {P.}~\bibnamefont {Gogna}},\ }\href {\doibase
  https://doi.org/10.1002/adma.200600527} {\bibfield  {journal} {\bibinfo
  {journal} {Adv. Mater.}\ }\textbf {\bibinfo {volume} {19}},\ \bibinfo {pages}
  {1043} (\bibinfo {year} {2007})}\BibitemShut {NoStop}%
\bibitem [{\citenamefont {Alam}\ and\ \citenamefont
  {Ramakrishna}(2013)}]{alam2013review}%
  \BibitemOpen
  \bibfield  {author} {\bibinfo {author} {\bibfnamefont {H.}~\bibnamefont
  {Alam}}\ and\ \bibinfo {author} {\bibfnamefont {S.}~\bibnamefont
  {Ramakrishna}},\ }\href {\doibase
  https://doi.org/10.1016/j.nanoen.2012.10.005} {\bibfield  {journal} {\bibinfo
   {journal} {Nano Energy}\ }\textbf {\bibinfo {volume} {2}},\ \bibinfo {pages}
  {190} (\bibinfo {year} {2013})}\BibitemShut {NoStop}%
\bibitem [{\citenamefont {Novoselov}\ \emph {et~al.}(2004)\citenamefont
  {Novoselov}, \citenamefont {Geim}, \citenamefont {Morozov}, \citenamefont
  {Jiang}, \citenamefont {Zhang}, \citenamefont {Dubonos}, \citenamefont
  {Grigorieva},\ and\ \citenamefont {Firsov}}]{novoselov2004electric}%
  \BibitemOpen
  \bibfield  {author} {\bibinfo {author} {\bibfnamefont {K.~S.}\ \bibnamefont
  {Novoselov}}, \bibinfo {author} {\bibfnamefont {A.~K.}\ \bibnamefont {Geim}},
  \bibinfo {author} {\bibfnamefont {S.~V.}\ \bibnamefont {Morozov}}, \bibinfo
  {author} {\bibfnamefont {D.}~\bibnamefont {Jiang}}, \bibinfo {author}
  {\bibfnamefont {Y.}~\bibnamefont {Zhang}}, \bibinfo {author} {\bibfnamefont
  {S.~V.}\ \bibnamefont {Dubonos}}, \bibinfo {author} {\bibfnamefont {I.~V.}\
  \bibnamefont {Grigorieva}}, \ and\ \bibinfo {author} {\bibfnamefont {A.~A.}\
  \bibnamefont {Firsov}},\ }\href {\doibase 10.1126/science.1102896} {\bibfield
   {journal} {\bibinfo  {journal} {Science}\ }\textbf {\bibinfo {volume}
  {306}},\ \bibinfo {pages} {666} (\bibinfo {year} {2004})}\BibitemShut
  {NoStop}%
\bibitem [{\citenamefont {Li}\ \emph {et~al.}(2020)\citenamefont {Li},
  \citenamefont {Gong}, \citenamefont {Chen}, \citenamefont {Lin},
  \citenamefont {Khan}, \citenamefont {Zhang}, \citenamefont {Li},
  \citenamefont {Zhang},\ and\ \citenamefont {Xie}}]{li2020recent}%
  \BibitemOpen
  \bibfield  {author} {\bibinfo {author} {\bibfnamefont {D.}~\bibnamefont
  {Li}}, \bibinfo {author} {\bibfnamefont {Y.}~\bibnamefont {Gong}}, \bibinfo
  {author} {\bibfnamefont {Y.}~\bibnamefont {Chen}}, \bibinfo {author}
  {\bibfnamefont {J.}~\bibnamefont {Lin}}, \bibinfo {author} {\bibfnamefont
  {Q.}~\bibnamefont {Khan}}, \bibinfo {author} {\bibfnamefont {Y.}~\bibnamefont
  {Zhang}}, \bibinfo {author} {\bibfnamefont {Y.}~\bibnamefont {Li}}, \bibinfo
  {author} {\bibfnamefont {H.}~\bibnamefont {Zhang}}, \ and\ \bibinfo {author}
  {\bibfnamefont {H.}~\bibnamefont {Xie}},\ }\href {\doibase
  10.1007/s40820-020-0374-x} {\bibfield  {journal} {\bibinfo  {journal}
  {Nanomicro Lett}\ }\textbf {\bibinfo {volume} {12}},\ \bibinfo {pages} {1}
  (\bibinfo {year} {2020})}\BibitemShut {NoStop}%
\bibitem [{\citenamefont {Naguib}\ \emph {et~al.}(2011)\citenamefont {Naguib},
  \citenamefont {Kurtoglu}, \citenamefont {Presser}, \citenamefont {Lu},
  \citenamefont {Niu}, \citenamefont {Heon}, \citenamefont {Hultman},
  \citenamefont {Gogotsi},\ and\ \citenamefont {Barsoum}}]{naguib2011two}%
  \BibitemOpen
  \bibfield  {author} {\bibinfo {author} {\bibfnamefont {M.}~\bibnamefont
  {Naguib}}, \bibinfo {author} {\bibfnamefont {M.}~\bibnamefont {Kurtoglu}},
  \bibinfo {author} {\bibfnamefont {V.}~\bibnamefont {Presser}}, \bibinfo
  {author} {\bibfnamefont {J.}~\bibnamefont {Lu}}, \bibinfo {author}
  {\bibfnamefont {J.}~\bibnamefont {Niu}}, \bibinfo {author} {\bibfnamefont
  {M.}~\bibnamefont {Heon}}, \bibinfo {author} {\bibfnamefont {L.}~\bibnamefont
  {Hultman}}, \bibinfo {author} {\bibfnamefont {Y.}~\bibnamefont {Gogotsi}}, \
  and\ \bibinfo {author} {\bibfnamefont {M.~W.}\ \bibnamefont {Barsoum}},\
  }\href {\doibase https://doi.org/10.1002/adma.201102306} {\bibfield
  {journal} {\bibinfo  {journal} {Adv. Mater.}\ }\textbf {\bibinfo {volume}
  {23}},\ \bibinfo {pages} {4248} (\bibinfo {year} {2011})}\BibitemShut
  {NoStop}%
\bibitem [{\citenamefont {Hu}\ \emph {et~al.}(2015)\citenamefont {Hu},
  \citenamefont {Lai}, \citenamefont {Tao}, \citenamefont {Lu}, \citenamefont
  {Halim}, \citenamefont {Sun}, \citenamefont {Zhang}, \citenamefont {Yang},
  \citenamefont {Anasori}, \citenamefont {Wang}, \citenamefont {Sakka},
  \citenamefont {Hultman}, \citenamefont {Eklund}, \citenamefont {Rosen},\ and\
  \citenamefont {Barsoum}}]{hu2015mo}%
  \BibitemOpen
  \bibfield  {author} {\bibinfo {author} {\bibfnamefont {C.}~\bibnamefont
  {Hu}}, \bibinfo {author} {\bibfnamefont {C.-C.}\ \bibnamefont {Lai}},
  \bibinfo {author} {\bibfnamefont {Q.}~\bibnamefont {Tao}}, \bibinfo {author}
  {\bibfnamefont {J.}~\bibnamefont {Lu}}, \bibinfo {author} {\bibfnamefont
  {J.}~\bibnamefont {Halim}}, \bibinfo {author} {\bibfnamefont
  {L.}~\bibnamefont {Sun}}, \bibinfo {author} {\bibfnamefont {J.}~\bibnamefont
  {Zhang}}, \bibinfo {author} {\bibfnamefont {J.}~\bibnamefont {Yang}},
  \bibinfo {author} {\bibfnamefont {B.}~\bibnamefont {Anasori}}, \bibinfo
  {author} {\bibfnamefont {J.}~\bibnamefont {Wang}}, \bibinfo {author}
  {\bibfnamefont {Y.}~\bibnamefont {Sakka}}, \bibinfo {author} {\bibfnamefont
  {L.}~\bibnamefont {Hultman}}, \bibinfo {author} {\bibfnamefont
  {P.}~\bibnamefont {Eklund}}, \bibinfo {author} {\bibfnamefont
  {J.}~\bibnamefont {Rosen}}, \ and\ \bibinfo {author} {\bibfnamefont {M.~W.}\
  \bibnamefont {Barsoum}},\ }\href {\doibase 10.1039/C5CC00980D} {\bibfield
  {journal} {\bibinfo  {journal} {Chem. Commun.}\ }\textbf {\bibinfo {volume}
  {51}},\ \bibinfo {pages} {6560} (\bibinfo {year} {2015})}\BibitemShut
  {NoStop}%
\bibitem [{\citenamefont {Lukatskaya}\ \emph {et~al.}(2013)\citenamefont
  {Lukatskaya}, \citenamefont {Mashtalir}, \citenamefont {Ren}, \citenamefont
  {Dall’Agnese}, \citenamefont {Rozier}, \citenamefont {Taberna},
  \citenamefont {Naguib}, \citenamefont {Simon}, \citenamefont {Barsoum},\ and\
  \citenamefont {Gogotsi}}]{lukatskaya2013cation}%
  \BibitemOpen
  \bibfield  {author} {\bibinfo {author} {\bibfnamefont {M.~R.}\ \bibnamefont
  {Lukatskaya}}, \bibinfo {author} {\bibfnamefont {O.}~\bibnamefont
  {Mashtalir}}, \bibinfo {author} {\bibfnamefont {C.~E.}\ \bibnamefont {Ren}},
  \bibinfo {author} {\bibfnamefont {Y.}~\bibnamefont {Dall’Agnese}}, \bibinfo
  {author} {\bibfnamefont {P.}~\bibnamefont {Rozier}}, \bibinfo {author}
  {\bibfnamefont {P.~L.}\ \bibnamefont {Taberna}}, \bibinfo {author}
  {\bibfnamefont {M.}~\bibnamefont {Naguib}}, \bibinfo {author} {\bibfnamefont
  {P.}~\bibnamefont {Simon}}, \bibinfo {author} {\bibfnamefont {M.~W.}\
  \bibnamefont {Barsoum}}, \ and\ \bibinfo {author} {\bibfnamefont
  {Y.}~\bibnamefont {Gogotsi}},\ }\href {\doibase 10.1126/science.1241488}
  {\bibfield  {journal} {\bibinfo  {journal} {Science}\ }\textbf {\bibinfo
  {volume} {341}},\ \bibinfo {pages} {1502} (\bibinfo {year}
  {2013})}\BibitemShut {NoStop}%
\bibitem [{\citenamefont {Xie}\ \emph {et~al.}(2013)\citenamefont {Xie},
  \citenamefont {Chen}, \citenamefont {Ding}, \citenamefont {Nie},\ and\
  \citenamefont {Wei}}]{xie2013extraordinarily}%
  \BibitemOpen
  \bibfield  {author} {\bibinfo {author} {\bibfnamefont {X.}~\bibnamefont
  {Xie}}, \bibinfo {author} {\bibfnamefont {S.}~\bibnamefont {Chen}}, \bibinfo
  {author} {\bibfnamefont {W.}~\bibnamefont {Ding}}, \bibinfo {author}
  {\bibfnamefont {Y.}~\bibnamefont {Nie}}, \ and\ \bibinfo {author}
  {\bibfnamefont {Z.}~\bibnamefont {Wei}},\ }\href {\doibase
  10.1039/C3CC44428G} {\bibfield  {journal} {\bibinfo  {journal} {Chem.
  Commun.}\ }\textbf {\bibinfo {volume} {49}},\ \bibinfo {pages} {10112}
  (\bibinfo {year} {2013})}\BibitemShut {NoStop}%
\bibitem [{\citenamefont {Shahzad}\ \emph {et~al.}(2016)\citenamefont
  {Shahzad}, \citenamefont {Alhabeb}, \citenamefont {Hatter}, \citenamefont
  {Anasori}, \citenamefont {Hong}, \citenamefont {Koo},\ and\ \citenamefont
  {Gogotsi}}]{shahzad2016electromagnetic}%
  \BibitemOpen
  \bibfield  {author} {\bibinfo {author} {\bibfnamefont {F.}~\bibnamefont
  {Shahzad}}, \bibinfo {author} {\bibfnamefont {M.}~\bibnamefont {Alhabeb}},
  \bibinfo {author} {\bibfnamefont {C.~B.}\ \bibnamefont {Hatter}}, \bibinfo
  {author} {\bibfnamefont {B.}~\bibnamefont {Anasori}}, \bibinfo {author}
  {\bibfnamefont {S.~M.}\ \bibnamefont {Hong}}, \bibinfo {author}
  {\bibfnamefont {C.~M.}\ \bibnamefont {Koo}}, \ and\ \bibinfo {author}
  {\bibfnamefont {Y.}~\bibnamefont {Gogotsi}},\ }\href {\doibase
  10.1126/science.aag2421} {\bibfield  {journal} {\bibinfo  {journal}
  {Science}\ }\textbf {\bibinfo {volume} {353}},\ \bibinfo {pages} {1137}
  (\bibinfo {year} {2016})}\BibitemShut {NoStop}%
\bibitem [{\citenamefont {Liu}\ \emph {et~al.}(2020)\citenamefont {Liu},
  \citenamefont {Ding}, \citenamefont {Liu}, \citenamefont {Shen},
  \citenamefont {Jiang}, \citenamefont {Liu}, \citenamefont {Zhu},
  \citenamefont {Zhang}, \citenamefont {Liu},\ and\ \citenamefont
  {Xu}}]{filmti3c2}%
  \BibitemOpen
  \bibfield  {author} {\bibinfo {author} {\bibfnamefont {P.}~\bibnamefont
  {Liu}}, \bibinfo {author} {\bibfnamefont {W.}~\bibnamefont {Ding}}, \bibinfo
  {author} {\bibfnamefont {J.}~\bibnamefont {Liu}}, \bibinfo {author}
  {\bibfnamefont {L.}~\bibnamefont {Shen}}, \bibinfo {author} {\bibfnamefont
  {F.}~\bibnamefont {Jiang}}, \bibinfo {author} {\bibfnamefont
  {P.}~\bibnamefont {Liu}}, \bibinfo {author} {\bibfnamefont {Z.}~\bibnamefont
  {Zhu}}, \bibinfo {author} {\bibfnamefont {G.}~\bibnamefont {Zhang}}, \bibinfo
  {author} {\bibfnamefont {C.}~\bibnamefont {Liu}}, \ and\ \bibinfo {author}
  {\bibfnamefont {J.}~\bibnamefont {Xu}},\ }\href {\doibase
  https://doi.org/10.1016/j.jallcom.2020.154634} {\bibfield  {journal}
  {\bibinfo  {journal} {J. Alloys and Compd.}\ }\textbf {\bibinfo {volume}
  {829}},\ \bibinfo {pages} {154634} (\bibinfo {year} {2020})}\BibitemShut
  {NoStop}%
\bibitem [{\citenamefont {Jhon}\ \emph {et~al.}(2018)\citenamefont {Jhon},
  \citenamefont {Seo},\ and\ \citenamefont {Jhon}}]{flaketi3c2}%
  \BibitemOpen
  \bibfield  {author} {\bibinfo {author} {\bibfnamefont {Y.~I.}\ \bibnamefont
  {Jhon}}, \bibinfo {author} {\bibfnamefont {M.}~\bibnamefont {Seo}}, \ and\
  \bibinfo {author} {\bibfnamefont {Y.~M.}\ \bibnamefont {Jhon}},\ }\href
  {\doibase 10.1039/C7NR05351G} {\bibfield  {journal} {\bibinfo  {journal}
  {Nanoscale}\ }\textbf {\bibinfo {volume} {10}},\ \bibinfo {pages} {69}
  (\bibinfo {year} {2018})}\BibitemShut {NoStop}%
\bibitem [{\citenamefont {Kim}\ \emph {et~al.}(2017)\citenamefont {Kim},
  \citenamefont {Anasori}, \citenamefont {Gogotsi},\ and\ \citenamefont
  {Alshareef}}]{kim2017thermoelectric}%
  \BibitemOpen
  \bibfield  {author} {\bibinfo {author} {\bibfnamefont {H.}~\bibnamefont
  {Kim}}, \bibinfo {author} {\bibfnamefont {B.}~\bibnamefont {Anasori}},
  \bibinfo {author} {\bibfnamefont {Y.}~\bibnamefont {Gogotsi}}, \ and\
  \bibinfo {author} {\bibfnamefont {H.~N.}\ \bibnamefont {Alshareef}},\ }\href
  {\doibase 10.1021/acs.chemmater.7b02056} {\bibfield  {journal} {\bibinfo
  {journal} {Chem. Mater.}\ }\textbf {\bibinfo {volume} {29}},\ \bibinfo
  {pages} {6472} (\bibinfo {year} {2017})}\BibitemShut {NoStop}%
\bibitem [{\citenamefont {Huang}\ \emph {et~al.}(2022)\citenamefont {Huang},
  \citenamefont {Kim}, \citenamefont {Zou}, \citenamefont {Xu}, \citenamefont
  {Zhu}, \citenamefont {Ahmad}, \citenamefont {Almutairi},\ and\ \citenamefont
  {Alshareef}}]{nanogen}%
  \BibitemOpen
  \bibfield  {author} {\bibinfo {author} {\bibfnamefont {D.}~\bibnamefont
  {Huang}}, \bibinfo {author} {\bibfnamefont {H.}~\bibnamefont {Kim}}, \bibinfo
  {author} {\bibfnamefont {G.}~\bibnamefont {Zou}}, \bibinfo {author}
  {\bibfnamefont {X.}~\bibnamefont {Xu}}, \bibinfo {author} {\bibfnamefont
  {Y.}~\bibnamefont {Zhu}}, \bibinfo {author} {\bibfnamefont {K.}~\bibnamefont
  {Ahmad}}, \bibinfo {author} {\bibfnamefont {Z.~A.}\ \bibnamefont
  {Almutairi}}, \ and\ \bibinfo {author} {\bibfnamefont {H.~N.}\ \bibnamefont
  {Alshareef}},\ }\href {\doibase https://doi.org/10.1016/j.mtener.2022.101129}
  {\bibfield  {journal} {\bibinfo  {journal} {Mater. Today Energy}\ }\textbf
  {\bibinfo {volume} {29}},\ \bibinfo {pages} {101129} (\bibinfo {year}
  {2022})}\BibitemShut {NoStop}%
\bibitem [{\citenamefont {Kumar}\ and\ \citenamefont
  {Schwingenschl\"ogl}(2016)}]{sc2c}%
  \BibitemOpen
  \bibfield  {author} {\bibinfo {author} {\bibfnamefont {S.}~\bibnamefont
  {Kumar}}\ and\ \bibinfo {author} {\bibfnamefont {U.}~\bibnamefont
  {Schwingenschl\"ogl}},\ }\href {\doibase 10.1103/PhysRevB.94.035405}
  {\bibfield  {journal} {\bibinfo  {journal} {Phys. Rev. B}\ }\textbf {\bibinfo
  {volume} {94}},\ \bibinfo {pages} {035405} (\bibinfo {year}
  {2016})}\BibitemShut {NoStop}%
\bibitem [{\citenamefont {Guo}\ \emph {et~al.}(2018)\citenamefont {Guo},
  \citenamefont {Miao}, \citenamefont {Zhou}, \citenamefont {Pan},\ and\
  \citenamefont {Sun}}]{pccp2018}%
  \BibitemOpen
  \bibfield  {author} {\bibinfo {author} {\bibfnamefont {Z.}~\bibnamefont
  {Guo}}, \bibinfo {author} {\bibfnamefont {N.}~\bibnamefont {Miao}}, \bibinfo
  {author} {\bibfnamefont {J.}~\bibnamefont {Zhou}}, \bibinfo {author}
  {\bibfnamefont {Y.}~\bibnamefont {Pan}}, \ and\ \bibinfo {author}
  {\bibfnamefont {Z.}~\bibnamefont {Sun}},\ }\href {\doibase
  10.1039/C8CP02564A} {\bibfield  {journal} {\bibinfo  {journal} {Phys. Chem.
  Chem. Phys.}\ }\textbf {\bibinfo {volume} {20}},\ \bibinfo {pages} {19689}
  (\bibinfo {year} {2018})}\BibitemShut {NoStop}%
\bibitem [{\citenamefont {Omugbe}\ \emph {et~al.}(2022)\citenamefont {Omugbe},
  \citenamefont {Osafile}, \citenamefont {Nenuwe},\ and\ \citenamefont
  {Enaibe}}]{y2ct2}%
  \BibitemOpen
  \bibfield  {author} {\bibinfo {author} {\bibfnamefont {E.}~\bibnamefont
  {Omugbe}}, \bibinfo {author} {\bibfnamefont {O.}~\bibnamefont {Osafile}},
  \bibinfo {author} {\bibfnamefont {O.}~\bibnamefont {Nenuwe}}, \ and\ \bibinfo
  {author} {\bibfnamefont {E.}~\bibnamefont {Enaibe}},\ }\href {\doibase
  https://doi.org/10.1016/j.physb.2022.413922} {\bibfield  {journal} {\bibinfo
  {journal} {Physica B Condens. Matter}\ }\textbf {\bibinfo {volume} {639}},\
  \bibinfo {pages} {413922} (\bibinfo {year} {2022})}\BibitemShut {NoStop}%
\bibitem [{\citenamefont {Gandi}\ \emph {et~al.}(2016)\citenamefont {Gandi},
  \citenamefont {Alshareef},\ and\ \citenamefont
  {Schwingenschlögl}}]{gandi2016thermoelectric}%
  \BibitemOpen
  \bibfield  {author} {\bibinfo {author} {\bibfnamefont {A.~N.}\ \bibnamefont
  {Gandi}}, \bibinfo {author} {\bibfnamefont {H.~N.}\ \bibnamefont
  {Alshareef}}, \ and\ \bibinfo {author} {\bibfnamefont {U.}~\bibnamefont
  {Schwingenschlögl}},\ }\href {\doibase 10.1021/acs.chemmater.5b04257}
  {\bibfield  {journal} {\bibinfo  {journal} {Chem. Mater.}\ }\textbf {\bibinfo
  {volume} {28}},\ \bibinfo {pages} {1647} (\bibinfo {year}
  {2016})}\BibitemShut {NoStop}%
\bibitem [{\citenamefont {Sarikurt}\ \emph {et~al.}(2018)\citenamefont
  {Sarikurt}, \citenamefont {Çakır}, \citenamefont {Keçeli},\ and\
  \citenamefont {Sevik}}]{sarikurt2018influence}%
  \BibitemOpen
  \bibfield  {author} {\bibinfo {author} {\bibfnamefont {S.}~\bibnamefont
  {Sarikurt}}, \bibinfo {author} {\bibfnamefont {D.}~\bibnamefont {Çakır}},
  \bibinfo {author} {\bibfnamefont {M.}~\bibnamefont {Keçeli}}, \ and\
  \bibinfo {author} {\bibfnamefont {C.}~\bibnamefont {Sevik}},\ }\href
  {\doibase 10.1039/C7NR09144C} {\bibfield  {journal} {\bibinfo  {journal}
  {Nanoscale}\ }\textbf {\bibinfo {volume} {10}},\ \bibinfo {pages} {8859}
  (\bibinfo {year} {2018})}\BibitemShut {NoStop}%
\bibitem [{\citenamefont {Karmakar}\ and\ \citenamefont
  {Saha-Dasgupta}(2020)}]{tanusri}%
  \BibitemOpen
  \bibfield  {author} {\bibinfo {author} {\bibfnamefont {S.}~\bibnamefont
  {Karmakar}}\ and\ \bibinfo {author} {\bibfnamefont {T.}~\bibnamefont
  {Saha-Dasgupta}},\ }\href {\doibase 10.1103/PhysRevMaterials.4.124007}
  {\bibfield  {journal} {\bibinfo  {journal} {Phys. Rev. Mater.}\ }\textbf
  {\bibinfo {volume} {4}},\ \bibinfo {pages} {124007} (\bibinfo {year}
  {2020})}\BibitemShut {NoStop}%
\bibitem [{\citenamefont {Jing}\ \emph {et~al.}(2019)\citenamefont {Jing},
  \citenamefont {Wang}, \citenamefont {Feng}, \citenamefont {Xiao},
  \citenamefont {Ding}, \citenamefont {Wu},\ and\ \citenamefont
  {Cheng}}]{jing2019superior}%
  \BibitemOpen
  \bibfield  {author} {\bibinfo {author} {\bibfnamefont {Z.}~\bibnamefont
  {Jing}}, \bibinfo {author} {\bibfnamefont {H.}~\bibnamefont {Wang}}, \bibinfo
  {author} {\bibfnamefont {X.}~\bibnamefont {Feng}}, \bibinfo {author}
  {\bibfnamefont {B.}~\bibnamefont {Xiao}}, \bibinfo {author} {\bibfnamefont
  {Y.}~\bibnamefont {Ding}}, \bibinfo {author} {\bibfnamefont {K.}~\bibnamefont
  {Wu}}, \ and\ \bibinfo {author} {\bibfnamefont {Y.}~\bibnamefont {Cheng}},\
  }\href {\doibase 10.1021/acs.jpclett.9b01827} {\bibfield  {journal} {\bibinfo
   {journal} {J. Phys. Chem. Lett.}\ }\textbf {\bibinfo {volume} {10}},\
  \bibinfo {pages} {5721} (\bibinfo {year} {2019})}\BibitemShut {NoStop}%
\bibitem [{\citenamefont {Lu}\ \emph {et~al.}(2017)\citenamefont {Lu},
  \citenamefont {Zhu}, \citenamefont {Xiao}, \citenamefont {Chuu},
  \citenamefont {Han}, \citenamefont {Chiu}, \citenamefont {Cheng},
  \citenamefont {Yang}, \citenamefont {Wei}, \citenamefont {Yang} \emph
  {et~al.}}]{lu2017janus}%
  \BibitemOpen
  \bibfield  {author} {\bibinfo {author} {\bibfnamefont {A.-Y.}\ \bibnamefont
  {Lu}}, \bibinfo {author} {\bibfnamefont {H.}~\bibnamefont {Zhu}}, \bibinfo
  {author} {\bibfnamefont {J.}~\bibnamefont {Xiao}}, \bibinfo {author}
  {\bibfnamefont {C.-P.}\ \bibnamefont {Chuu}}, \bibinfo {author}
  {\bibfnamefont {Y.}~\bibnamefont {Han}}, \bibinfo {author} {\bibfnamefont
  {M.-H.}\ \bibnamefont {Chiu}}, \bibinfo {author} {\bibfnamefont {C.-C.}\
  \bibnamefont {Cheng}}, \bibinfo {author} {\bibfnamefont {C.-W.}\ \bibnamefont
  {Yang}}, \bibinfo {author} {\bibfnamefont {K.-H.}\ \bibnamefont {Wei}},
  \bibinfo {author} {\bibfnamefont {Y.}~\bibnamefont {Yang}},  \emph {et~al.},\
  }\href {\doibase 10.1038/nnano.2017.100} {\bibfield  {journal} {\bibinfo
  {journal} {Nat. Nanotechnol.}\ }\textbf {\bibinfo {volume} {12}},\ \bibinfo
  {pages} {744} (\bibinfo {year} {2017})}\BibitemShut {NoStop}%
\bibitem [{\citenamefont {Deng}\ \emph {et~al.}(2019)\citenamefont {Deng},
  \citenamefont {Li}, \citenamefont {Guy},\ and\ \citenamefont
  {Zhang}}]{deng2019enhanced}%
  \BibitemOpen
  \bibfield  {author} {\bibinfo {author} {\bibfnamefont {S.}~\bibnamefont
  {Deng}}, \bibinfo {author} {\bibfnamefont {L.}~\bibnamefont {Li}}, \bibinfo
  {author} {\bibfnamefont {O.~J.}\ \bibnamefont {Guy}}, \ and\ \bibinfo
  {author} {\bibfnamefont {Y.}~\bibnamefont {Zhang}},\ }\href {\doibase
  10.1039/C9CP03639C} {\bibfield  {journal} {\bibinfo  {journal} {Phys. Chem.
  Chem. Phys.}\ }\textbf {\bibinfo {volume} {21}},\ \bibinfo {pages} {18161}
  (\bibinfo {year} {2019})}\BibitemShut {NoStop}%
\bibitem [{\citenamefont {Patel}\ \emph {et~al.}(2020)\citenamefont {Patel},
  \citenamefont {Singh}, \citenamefont {Sonvane}, \citenamefont {Thakor},\ and\
  \citenamefont {Ahuja}}]{patel2020high}%
  \BibitemOpen
  \bibfield  {author} {\bibinfo {author} {\bibfnamefont {A.}~\bibnamefont
  {Patel}}, \bibinfo {author} {\bibfnamefont {D.}~\bibnamefont {Singh}},
  \bibinfo {author} {\bibfnamefont {Y.}~\bibnamefont {Sonvane}}, \bibinfo
  {author} {\bibfnamefont {P.~B.}\ \bibnamefont {Thakor}}, \ and\ \bibinfo
  {author} {\bibfnamefont {R.}~\bibnamefont {Ahuja}},\ }\href {\doibase
  10.1021/acsami.0c13960} {\bibfield  {journal} {\bibinfo  {journal} {ACS Appl.
  Mater. Interfaces}\ }\textbf {\bibinfo {volume} {12}},\ \bibinfo {pages}
  {46212} (\bibinfo {year} {2020})}\BibitemShut {NoStop}%
\bibitem [{\citenamefont {Wong}\ \emph {et~al.}(2020)\citenamefont {Wong},
  \citenamefont {Deng}, \citenamefont {Shi}, \citenamefont {Wu}, \citenamefont
  {Tan},\ and\ \citenamefont {Yang}}]{wong2020high}%
  \BibitemOpen
  \bibfield  {author} {\bibinfo {author} {\bibfnamefont {Z.~M.}\ \bibnamefont
  {Wong}}, \bibinfo {author} {\bibfnamefont {T.}~\bibnamefont {Deng}}, \bibinfo
  {author} {\bibfnamefont {W.}~\bibnamefont {Shi}}, \bibinfo {author}
  {\bibfnamefont {G.}~\bibnamefont {Wu}}, \bibinfo {author} {\bibfnamefont
  {T.~L.}\ \bibnamefont {Tan}}, \ and\ \bibinfo {author} {\bibfnamefont
  {S.-W.}\ \bibnamefont {Yang}},\ }\href {\doibase 10.1039/D0MA00391C}
  {\bibfield  {journal} {\bibinfo  {journal} {Mater. Adv.}\ }\textbf {\bibinfo
  {volume} {1}},\ \bibinfo {pages} {1176} (\bibinfo {year} {2020})}\BibitemShut
  {NoStop}%
\bibitem [{\citenamefont {Lu}\ \emph {et~al.}(2022)\citenamefont {Lu},
  \citenamefont {Ren}, \citenamefont {He}, \citenamefont {Yu}, \citenamefont
  {Jiang},\ and\ \citenamefont {Chen}}]{ta2cs2}%
  \BibitemOpen
  \bibfield  {author} {\bibinfo {author} {\bibfnamefont {S.}~\bibnamefont
  {Lu}}, \bibinfo {author} {\bibfnamefont {W.}~\bibnamefont {Ren}}, \bibinfo
  {author} {\bibfnamefont {J.}~\bibnamefont {He}}, \bibinfo {author}
  {\bibfnamefont {C.}~\bibnamefont {Yu}}, \bibinfo {author} {\bibfnamefont
  {P.}~\bibnamefont {Jiang}}, \ and\ \bibinfo {author} {\bibfnamefont
  {J.}~\bibnamefont {Chen}},\ }\href {\doibase 10.1103/PhysRevB.105.165301}
  {\bibfield  {journal} {\bibinfo  {journal} {Phys. Rev. B}\ }\textbf {\bibinfo
  {volume} {105}},\ \bibinfo {pages} {165301} (\bibinfo {year}
  {2022})}\BibitemShut {NoStop}%
\bibitem [{\citenamefont {Hohenberg}\ and\ \citenamefont
  {Kohn}(1964)}]{hohenberg1964inhomogeneous}%
  \BibitemOpen
  \bibfield  {author} {\bibinfo {author} {\bibfnamefont {P.}~\bibnamefont
  {Hohenberg}}\ and\ \bibinfo {author} {\bibfnamefont {W.}~\bibnamefont
  {Kohn}},\ }\href {\doibase 10.1103/PhysRev.136.B864} {\bibfield  {journal}
  {\bibinfo  {journal} {Phys. Rev.}\ }\textbf {\bibinfo {volume} {136}},\
  \bibinfo {pages} {B864} (\bibinfo {year} {1964})}\BibitemShut {NoStop}%
\bibitem [{\citenamefont {Kohn}\ and\ \citenamefont
  {Sham}(1965)}]{kohn1965self}%
  \BibitemOpen
  \bibfield  {author} {\bibinfo {author} {\bibfnamefont {W.}~\bibnamefont
  {Kohn}}\ and\ \bibinfo {author} {\bibfnamefont {L.~J.}\ \bibnamefont
  {Sham}},\ }\href {\doibase 10.1103/PhysRev.140.A1133} {\bibfield  {journal}
  {\bibinfo  {journal} {Phys. Rev.}\ }\textbf {\bibinfo {volume} {140}},\
  \bibinfo {pages} {A1133} (\bibinfo {year} {1965})}\BibitemShut {NoStop}%
\bibitem [{\citenamefont {Kresse}\ and\ \citenamefont
  {Furthm\"uller}(1996)}]{kresse1996efficient}%
  \BibitemOpen
  \bibfield  {author} {\bibinfo {author} {\bibfnamefont {G.}~\bibnamefont
  {Kresse}}\ and\ \bibinfo {author} {\bibfnamefont {J.}~\bibnamefont
  {Furthm\"uller}},\ }\href {\doibase 10.1103/PhysRevB.54.11169} {\bibfield
  {journal} {\bibinfo  {journal} {Phys. Rev. B}\ }\textbf {\bibinfo {volume}
  {54}},\ \bibinfo {pages} {11169} (\bibinfo {year} {1996})}\BibitemShut
  {NoStop}%
\bibitem [{\citenamefont {Bl\"ochl}(1994)}]{blochl1994projector}%
  \BibitemOpen
  \bibfield  {author} {\bibinfo {author} {\bibfnamefont {P.~E.}\ \bibnamefont
  {Bl\"ochl}},\ }\href {\doibase 10.1103/PhysRevB.50.17953} {\bibfield
  {journal} {\bibinfo  {journal} {Phys. Rev. B}\ }\textbf {\bibinfo {volume}
  {50}},\ \bibinfo {pages} {17953} (\bibinfo {year} {1994})}\BibitemShut
  {NoStop}%
\bibitem [{\citenamefont {Perdew}\ \emph {et~al.}(1996)\citenamefont {Perdew},
  \citenamefont {Burke},\ and\ \citenamefont
  {Ernzerhof}}]{perdew1996generalized}%
  \BibitemOpen
  \bibfield  {author} {\bibinfo {author} {\bibfnamefont {J.~P.}\ \bibnamefont
  {Perdew}}, \bibinfo {author} {\bibfnamefont {K.}~\bibnamefont {Burke}}, \
  and\ \bibinfo {author} {\bibfnamefont {M.}~\bibnamefont {Ernzerhof}},\ }\href
  {\doibase 10.1103/PhysRevLett.77.3865} {\bibfield  {journal} {\bibinfo
  {journal} {Phys. Rev. Lett.}\ }\textbf {\bibinfo {volume} {77}},\ \bibinfo
  {pages} {3865} (\bibinfo {year} {1996})}\BibitemShut {NoStop}%
\bibitem [{\citenamefont {Monkhorst}\ and\ \citenamefont
  {Pack}(1976)}]{monkhorst1976special}%
  \BibitemOpen
  \bibfield  {author} {\bibinfo {author} {\bibfnamefont {H.~J.}\ \bibnamefont
  {Monkhorst}}\ and\ \bibinfo {author} {\bibfnamefont {J.~D.}\ \bibnamefont
  {Pack}},\ }\href {\doibase 10.1103/PhysRevB.13.5188} {\bibfield  {journal}
  {\bibinfo  {journal} {Phys. Rev. B}\ }\textbf {\bibinfo {volume} {13}},\
  \bibinfo {pages} {5188} (\bibinfo {year} {1976})}\BibitemShut {NoStop}%
\bibitem [{\citenamefont {Nosé}(1984)}]{nose1984unified}%
  \BibitemOpen
  \bibfield  {author} {\bibinfo {author} {\bibfnamefont {S.}~\bibnamefont
  {Nosé}},\ }\href {\doibase 10.1063/1.447334} {\bibfield  {journal} {\bibinfo
   {journal} {J. Chem. Phys.}\ }\textbf {\bibinfo {volume} {81}},\ \bibinfo
  {pages} {511} (\bibinfo {year} {1984})}\BibitemShut {NoStop}%
\bibitem [{\citenamefont {Madsen}\ \emph {et~al.}(2018)\citenamefont {Madsen},
  \citenamefont {Carrete},\ and\ \citenamefont
  {Verstraete}}]{madsen2018boltztrap2}%
  \BibitemOpen
  \bibfield  {author} {\bibinfo {author} {\bibfnamefont {G.~K.}\ \bibnamefont
  {Madsen}}, \bibinfo {author} {\bibfnamefont {J.}~\bibnamefont {Carrete}}, \
  and\ \bibinfo {author} {\bibfnamefont {M.~J.}\ \bibnamefont {Verstraete}},\
  }\href {\doibase https://doi.org/10.1016/j.cpc.2018.05.010} {\bibfield
  {journal} {\bibinfo  {journal} {Comput. Phys. Commun.}\ }\textbf {\bibinfo
  {volume} {231}},\ \bibinfo {pages} {140} (\bibinfo {year}
  {2018})}\BibitemShut {NoStop}%
\bibitem [{\citenamefont {Bardeen}\ and\ \citenamefont
  {Shockley}(1950)}]{bardeen1950deformation}%
  \BibitemOpen
  \bibfield  {author} {\bibinfo {author} {\bibfnamefont {J.}~\bibnamefont
  {Bardeen}}\ and\ \bibinfo {author} {\bibfnamefont {W.}~\bibnamefont
  {Shockley}},\ }\href {\doibase 10.1103/PhysRev.80.72} {\bibfield  {journal}
  {\bibinfo  {journal} {Phys. Rev.}\ }\textbf {\bibinfo {volume} {80}},\
  \bibinfo {pages} {72} (\bibinfo {year} {1950})}\BibitemShut {NoStop}%
\bibitem [{\citenamefont {Togo}\ and\ \citenamefont
  {Tanaka}(2015)}]{togo2015first}%
  \BibitemOpen
  \bibfield  {author} {\bibinfo {author} {\bibfnamefont {A.}~\bibnamefont
  {Togo}}\ and\ \bibinfo {author} {\bibfnamefont {I.}~\bibnamefont {Tanaka}},\
  }\href {\doibase https://doi.org/10.1016/j.scriptamat.2015.07.021} {\bibfield
   {journal} {\bibinfo  {journal} {Scr. Mater.}\ }\textbf {\bibinfo {volume}
  {108}},\ \bibinfo {pages} {1} (\bibinfo {year} {2015})}\BibitemShut {NoStop}%
\bibitem [{\citenamefont {Li}\ \emph {et~al.}(2014)\citenamefont {Li},
  \citenamefont {Carrete}, \citenamefont {{A. Katcho}},\ and\ \citenamefont
  {Mingo}}]{li2014shengbte}%
  \BibitemOpen
  \bibfield  {author} {\bibinfo {author} {\bibfnamefont {W.}~\bibnamefont
  {Li}}, \bibinfo {author} {\bibfnamefont {J.}~\bibnamefont {Carrete}},
  \bibinfo {author} {\bibfnamefont {N.}~\bibnamefont {{A. Katcho}}}, \ and\
  \bibinfo {author} {\bibfnamefont {N.}~\bibnamefont {Mingo}},\ }\href
  {\doibase https://doi.org/10.1016/j.cpc.2014.02.015} {\bibfield  {journal}
  {\bibinfo  {journal} {Comput. Phys. Commun.}\ }\textbf {\bibinfo {volume}
  {185}},\ \bibinfo {pages} {1747} (\bibinfo {year} {2014})}\BibitemShut
  {NoStop}%
\bibitem [{\citenamefont {Khazaei}\ \emph {et~al.}(2014)\citenamefont
  {Khazaei}, \citenamefont {Arai}, \citenamefont {Sasaki}, \citenamefont
  {Estili},\ and\ \citenamefont {Sakka}}]{khazaei2014two}%
  \BibitemOpen
  \bibfield  {author} {\bibinfo {author} {\bibfnamefont {M.}~\bibnamefont
  {Khazaei}}, \bibinfo {author} {\bibfnamefont {M.}~\bibnamefont {Arai}},
  \bibinfo {author} {\bibfnamefont {T.}~\bibnamefont {Sasaki}}, \bibinfo
  {author} {\bibfnamefont {M.}~\bibnamefont {Estili}}, \ and\ \bibinfo {author}
  {\bibfnamefont {Y.}~\bibnamefont {Sakka}},\ }\href {\doibase
  10.1039/C4CP00467A} {\bibfield  {journal} {\bibinfo  {journal} {Phys. Chem.
  Chem. Phys.}\ }\textbf {\bibinfo {volume} {16}},\ \bibinfo {pages} {7841}
  (\bibinfo {year} {2014})}\BibitemShut {NoStop}%
\bibitem [{\citenamefont {Deringer}\ \emph {et~al.}(2011)\citenamefont
  {Deringer}, \citenamefont {Tchougréeff},\ and\ \citenamefont
  {Dronskowski}}]{deringer2011crystal}%
  \BibitemOpen
  \bibfield  {author} {\bibinfo {author} {\bibfnamefont {V.~L.}\ \bibnamefont
  {Deringer}}, \bibinfo {author} {\bibfnamefont {A.~L.}\ \bibnamefont
  {Tchougréeff}}, \ and\ \bibinfo {author} {\bibfnamefont {R.}~\bibnamefont
  {Dronskowski}},\ }\href {\doibase 10.1021/jp202489s} {\bibfield  {journal}
  {\bibinfo  {journal} {J. Phys. Chem. A}\ }\textbf {\bibinfo {volume} {115}},\
  \bibinfo {pages} {5461} (\bibinfo {year} {2011})}\BibitemShut {NoStop}%
\bibitem [{\citenamefont {Maintz}\ \emph {et~al.}(2016)\citenamefont {Maintz},
  \citenamefont {Deringer}, \citenamefont {Tchougréeff},\ and\ \citenamefont
  {Dronskowski}}]{maintz2016lobster}%
  \BibitemOpen
  \bibfield  {author} {\bibinfo {author} {\bibfnamefont {S.}~\bibnamefont
  {Maintz}}, \bibinfo {author} {\bibfnamefont {V.~L.}\ \bibnamefont
  {Deringer}}, \bibinfo {author} {\bibfnamefont {A.~L.}\ \bibnamefont
  {Tchougréeff}}, \ and\ \bibinfo {author} {\bibfnamefont {R.}~\bibnamefont
  {Dronskowski}},\ }\href {\doibase https://doi.org/10.1002/jcc.24300}
  {\bibfield  {journal} {\bibinfo  {journal} {J. Comput. Chem.}\ }\textbf
  {\bibinfo {volume} {37}},\ \bibinfo {pages} {1030} (\bibinfo {year}
  {2016})}\BibitemShut {NoStop}%
\bibitem [{\citenamefont {Shu}\ \emph {et~al.}(2023)\citenamefont {Shu},
  \citenamefont {Wang}, \citenamefont {Cui}, \citenamefont {Yan}, \citenamefont
  {Yan}, \citenamefont {Jia},\ and\ \citenamefont {Cai}}]{shu2023high}%
  \BibitemOpen
  \bibfield  {author} {\bibinfo {author} {\bibfnamefont {Z.}~\bibnamefont
  {Shu}}, \bibinfo {author} {\bibfnamefont {B.}~\bibnamefont {Wang}}, \bibinfo
  {author} {\bibfnamefont {X.}~\bibnamefont {Cui}}, \bibinfo {author}
  {\bibfnamefont {X.}~\bibnamefont {Yan}}, \bibinfo {author} {\bibfnamefont
  {H.}~\bibnamefont {Yan}}, \bibinfo {author} {\bibfnamefont {H.}~\bibnamefont
  {Jia}}, \ and\ \bibinfo {author} {\bibfnamefont {Y.}~\bibnamefont {Cai}},\
  }\href {\doibase https://doi.org/10.1016/j.cej.2022.140242} {\bibfield
  {journal} {\bibinfo  {journal} {Chem. Eng. J.}\ }\textbf {\bibinfo {volume}
  {454}},\ \bibinfo {pages} {140242} (\bibinfo {year} {2023})}\BibitemShut
  {NoStop}%
\bibitem [{\citenamefont {Slack}(1973)}]{slack1973nonmetallic}%
  \BibitemOpen
  \bibfield  {author} {\bibinfo {author} {\bibfnamefont {G.}~\bibnamefont
  {Slack}},\ }\href {\doibase https://doi.org/10.1016/0022-3697(73)90092-9}
  {\bibfield  {journal} {\bibinfo  {journal} {J. Phys. Chem. Solids}\ }\textbf
  {\bibinfo {volume} {34}},\ \bibinfo {pages} {321} (\bibinfo {year}
  {1973})}\BibitemShut {NoStop}%
\bibitem [{\citenamefont {Ziman}(2001)}]{ziman2001electrons}%
  \BibitemOpen
  \bibfield  {author} {\bibinfo {author} {\bibfnamefont {J.}~\bibnamefont
  {Ziman}},\ }\href {\doibase 10.1093/acprof:oso/9780198507796.001.0001} {\emph
  {\bibinfo {title} {{Electrons and Phonons: The Theory of Transport Phenomena
  in Solids}}}}\ (\bibinfo  {publisher} {Oxford University Press},\ \bibinfo
  {year} {2001})\BibitemShut {NoStop}%
\bibitem [{\citenamefont {Broido}\ \emph {et~al.}(2005)\citenamefont {Broido},
  \citenamefont {Ward},\ and\ \citenamefont {Mingo}}]{gruneisen}%
  \BibitemOpen
  \bibfield  {author} {\bibinfo {author} {\bibfnamefont {D.~A.}\ \bibnamefont
  {Broido}}, \bibinfo {author} {\bibfnamefont {A.}~\bibnamefont {Ward}}, \ and\
  \bibinfo {author} {\bibfnamefont {N.}~\bibnamefont {Mingo}},\ }\href
  {\doibase 10.1103/PhysRevB.72.014308} {\bibfield  {journal} {\bibinfo
  {journal} {Phys. Rev. B}\ }\textbf {\bibinfo {volume} {72}},\ \bibinfo
  {pages} {014308} (\bibinfo {year} {2005})}\BibitemShut {NoStop}%
\bibitem [{\citenamefont {Lindsay}\ and\ \citenamefont
  {Broido}(2008)}]{lindsay2008three}%
  \BibitemOpen
  \bibfield  {author} {\bibinfo {author} {\bibfnamefont {L.}~\bibnamefont
  {Lindsay}}\ and\ \bibinfo {author} {\bibfnamefont {D.~A.}\ \bibnamefont
  {Broido}},\ }\href {\doibase 10.1088/0953-8984/20/16/165209} {\bibfield
  {journal} {\bibinfo  {journal} {J. Phys.: Condens. Matter}\ }\textbf
  {\bibinfo {volume} {20}},\ \bibinfo {pages} {165209} (\bibinfo {year}
  {2008})}\BibitemShut {NoStop}%
\bibitem [{\citenamefont {Pandey}\ \emph {et~al.}(2017)\citenamefont {Pandey},
  \citenamefont {Polanco}, \citenamefont {Lindsay},\ and\ \citenamefont
  {Parker}}]{pandey2017lattice}%
  \BibitemOpen
  \bibfield  {author} {\bibinfo {author} {\bibfnamefont {T.}~\bibnamefont
  {Pandey}}, \bibinfo {author} {\bibfnamefont {C.~A.}\ \bibnamefont {Polanco}},
  \bibinfo {author} {\bibfnamefont {L.}~\bibnamefont {Lindsay}}, \ and\
  \bibinfo {author} {\bibfnamefont {D.~S.}\ \bibnamefont {Parker}},\ }\href
  {\doibase 10.1103/PhysRevB.95.224306} {\bibfield  {journal} {\bibinfo
  {journal} {Phys. Rev. B}\ }\textbf {\bibinfo {volume} {95}},\ \bibinfo
  {pages} {224306} (\bibinfo {year} {2017})}\BibitemShut {NoStop}%
\end{thebibliography}%

%
\end{document}